\documentclass[aps,amsmath,twocolumn,amssymb,floatfix,showpacs,superscriptaddress,nofootinbib,longbibliography]{revtex4-1}
\usepackage{braket}
\usepackage[dvipsnames]{xcolor}
\usepackage{float}
\usepackage{subfigure}
\usepackage{tikz}
\usepackage[colorlinks=true,linktoc=page,linkcolor=OliveGreen,citecolor=purple,urlcolor=violet]{hyperref}

\mathchardef\mhyphen="2D 

\newcommand{\ie}{{i.e.,\,\,}}

\newcommand\bea{\begin{eqnarray}}
\newcommand\eea{\end{eqnarray}}
\newcommand\beq{\begin{equation}}  
\newcommand\eeq{\end{equation}}

\newcommand{\non}{\nonumber}  
\usepackage[normalem]{ulem}
\definecolor{lime}{HTML}{A6CE39}
\usepackage{sidecap,tikz}
\DeclareRobustCommand{\orcidicon}{\hspace{-1.0mm}
	\begin{tikzpicture}
		\draw[lime, fill=lime] (0.0,0.0) 
		circle [radius=0.15] 
		node[white] {{\fontfamily{qag}\selectfont \tiny \,ID}};
		\draw[white, fill=white] (-0.0525,0.095) 
		circle [radius=0.007];
	\end{tikzpicture}
	\hspace{-3.0mm}
}
\foreach \x in {A, ..., Z}{\expandafter\xdef\csname orcid\x\endcsname{\noexpand\href{https://orcid.org/\csname orcidauthor\x\endcsname}{\noexpand\orcidicon}}
}

\AtBeginDocument{%
	\newwrite\bibnotes
	\def\bibnotesext{Notes.bib}
	\immediate\openout\bibnotes=\jobname\bibnotesext
	\immediate\write\bibnotes{@CONTROL{REVTEX41Control}}
	\immediate\write\bibnotes{@CONTROL{%
			apsrev41Control,author="08",editor="1",pages="1",title="1",year="1"}}
	\if@filesw
	\immediate\write\@auxout{\string\citation{apsrev41Control}}%
	\fi
}%

\begin{document}

\title{Quantum geometric signatures of Link-Unlink transitions and nonlinear Hall response in Hopf-link semimetals}  

\author{Kamalesh Bera\orcidA{}}
\email{kamalesh.bera@iopb.res.in}
\affiliation{Institute of Physics, Sachivalaya Marg, Bhubaneswar-751005, India}
\affiliation{Homi Bhabha National Institute, Training School Complex, Anushakti Nagar, Mumbai 400094, India}

\author{Arijit Saha\orcidC{}}
\email{arijit@iopb.res.in}
\affiliation{Institute of Physics, Sachivalaya Marg, Bhubaneswar-751005, India}
\affiliation{Homi Bhabha National Institute, Training School Complex, Anushakti Nagar, Mumbai 400094, India}

\author{Debashree Chowdhury\orcidB{}}
\email{debashreephys@gmail.com}
\affiliation{Centre for Nanotechnology, IIT Roorkee, Roorkee, Uttarakhand 247667, India}

\begin{abstract}
Quantum geometry, comprising of quantum metric and Berry curvature, plays a significant role in the electronic transport properties of solids. In this work, we theoretically investigate the quantum geometric properties of a Hopf-link semimetal, a distinct topological class that is charecterized by a nodal link-unlink transition. We compute the interband optical conductivity of the Hopf link, which effectively distinguishes between linked and trivial phases. While recent studies establish quantum metric dipole-mediated scattering-free nonlinear Hall effect, this effect becomes even more fascinating in systems where the Berry-curvature-dipole contribution to nonlinear Hall conductivity vanishes. Owing to the underlying $\mathcal{PT}$ symmetry of the Hopf-link semimetal, the Berry curvature and its corresponding contribution to the nonlinear Hall effect are entirely suppressed. Consequently, by introducing an appropriate perturbation, a finite nonlinear Hall conductivity emerges solely due to the quantum metric in the Hopf semimetal. Notably, this purely intrinsic, symmetry-driven nonlinear response remains entirely unmixed with extrinsic components.
\end{abstract}

\maketitle


\textcolor{blue}{\textit{Introduction:}-} The geometry of quantum states plays a pivotal role in modern condensed matter physics. One of its most prominent manifestations is the Berry phase~\cite{Berry_Phase}, a geometric phase that underlies a wide range of phenomena, including electric 
polarization~\cite{Thouless,Berry_Phase_RMP,Resta}, various Hall effects (anomalous, quantum, valley, and spin)~\cite{Hall_effects1,Hall_effects2,Hall_effects3,Hall_effects4,Hall_effects5,Hall_effects6}, orbital magnetism~\cite{Ob_mag1,Ob_mag2,Berry_Phase_RMP,Resta}, and quantum charge 
pumping~\cite{Berry_Phase_RMP}. However, a full description of quantum geometry demands an analysis beyond the Berry phase. A more comprehensive framework is offered by the quantum geometric tensor (QGT), which maps the local variation of quantum states in Hilbert space with crystal 
momentum~\cite{QGT_Provost,QGT_Paivi}. While the antisymmetric imaginary part of the QGT corresponds to the Berry curvature, the symmetric real part represents the quantum metric. In analogy to Berry curvature, the quantum metric drives a variety of remarkable physical phenomena, including superconductivity and superfluidity in flat-band 
systems~\cite{SF_SC_flat_band1,SF_SC_flat_band2,SF_SC_flat_band3,SF_SC_flat_band4}, quantum phase transitions~\cite{QPT}, bounds on topological band gaps~\cite{Bound_topological_gap}, 
current noise~\cite{Noise}, etc.
 
Furthermore, the study of the optical properties of electronic systems has emerged as a diagonostic tool to probe underlying electronic structure, yielding valuable insights into a wide range of materials, including Dirac and Weyl semimetals~\cite{Op_conduc_Dirac_Weyl1,Op_conduc_Dirac_Weyl2,Op_conduc_Dirac_Weyl3,Op_conduc_Dirac_Weyl4,Op_conduc_Dirac_Weyl5,Op_conduc_Dirac_Weyl6}, 
graphene~\cite{Op_graphene1,Op_graphene2}, and topological insulators~\cite{Op_TI}. Recent studies have established a bridge between optical conductivity and the quantum geometry of electronic states, demonstrating that optical measurements can serve as a direct probe of quantum-geometric 
properties~\cite{QGT_Op_conduct1,QGT_Op_conduct2,QGT_Op_conduct3,QGT_Op_conduct4}. Beyond linear response, nonlinear electrical transport has drawn significant attention owing to its intimate connection with band topology and quantum geometry.  In the nonlinear regime, transport properties are frequently explored via the Berry curvature dipole (BCD), which is evaluated from the momentum derivatives of the Berry curvature. However, the BCD captures only the extrinsic component of the nonlinear Hall conductivities (NLHCs), which is defined by its explicit dependence on the scattering time $\tau$. In contrast, uncovering the intrinsic, $\tau$-independent component of the NLHC requires accounting for interband transitions.  
This intrinsic NLHC can originate from the quantum metric dipole (QMD) of the electronic 
bands~\cite{NLHC_In4,NHLC_expt}.  In recent literature, both the extrinsic and intrinsic NLHC have been extensively investigated in a wide variety of 
materials~\cite{NLHC_Ex1,NLHC_Ex2,NLHC_Ex3,NLHC_Ex5,NLHC_Ex6,NLHC_In1,NLHC_In2,NLHC_In3,NLHC_In4,NLHC_In5,NLHC_In6}.

Over the past decade, topological semimetals have allured immense research interest due to their unconventional electronic properties and topological band structures. Based on the dimensionality of the band-crossing manifold between the conduction and valence bands, these systems are broadly classified into point-contact and line-contact semimetals. Weyl~\cite{Weyl_SM1,Weyl_SM2,Weyl_SM3,Weyl_SM4,Weyl_SM5,Weyl_SM6,Weyl_SM7,Weyl_SM8} and 
Dirac semimetals~\cite{Dirac_SM1,Dirac_SM2,Dirac_SM3,Dirac_SM4,Dirac_SM5} belong to the former category, whereas nodal-line 
semimetals~\cite{NLSM1,NLSM2,NLSM3,NLSM4,NLSM5,NLSM6,NLSM7} fit into the latter. In contrast to point-contact semimetals, nodal-line semimetals can host a richer variety of nodal structures beyond simple rings or straight lines~\cite{DTNL1,DTNL2,DTNL3,DTNL4}. The nodal lines can intersect either with themselves or with each other to form more intricate structures such as nodal chains, links, or knots. Prominent representatives of this behavior include nodal-knot and Hopf-link semimetals~\cite{Hopf1,Hopf2,Hopf3,Knot1,Knot2}. A Hopf-link semimetal is characterized by two nodal rings that are mutually linked, with each ring passing through the center of the other. By tuning an appropriate model parameter, the system undergoes a topological link-unlink transition, providing a unique platform for exploring the interplay between band topology and quantum geometry. The stability of this unique nodal structure usually depends on the symmetries associated with the system. Typically, these systems preserve parity-time reversal ($\mathcal{PT}$) and mirror symmetries.

Recent works~\cite{NLHC_In4,NLHC_In6,NLHC_In7} highlight that electronic systems featuring a purely intrinsic contribution to nonlinear Hall conductivity offer extreme experimental accessibility. This prioritizes the search for materials that manifest intrinsic contributions. In this letter, we pursue a twofold objective. First, we aim to identify the topological link-unlink transition by analyzing the features of optical conductivity. Second, we propose Hopf-link semimetals as a promising candidate to experimentally probe quantum-metric-mediated, purely intrinsic nonlinear Hall response. To begin with, we employ a low-energy continuum model to calculate the electronic band dispersion and demonstrate the topological link-unlink transition of the nodal rings. Then, we compute the $zz$-component of the quantum metric ($g_{zz}$) and interestingly find that, it exhibits distinct characteristic features across the link-unlink transition. Formulating the interband optical conductivity as a function of the quantum metric, we trace its evolution across this topological transition. Our analysis reveals that the linked phase exhibits distinct, sharp peaks in optical conductivity, which subsequently vanish in the unlinked phase. To elucidate the origin of these peaks, we compute the joint density of states (JDOS) and show that its peak positions match the characteristic frequencies of the interband optical conductivity. Furthermore, by examining the momentum-dependent band gap along one-dimensional momentum-space cuts to identify the locations responsible for these optical features, we establish that the peak intensities are directly correlated with the band gap. In the latter part of the work, we investigate the QMD-induced NLHC and show that the intrinsic NLHC 
is absent in the unperturbed Hopf-link semimetal. Conversely, introducing a symmetry-breaking perturbation lifts this restriction and generates a finite NLHC response. To uncover its microscopic origin, we analyze the momentum-space distribution of the QMD kernel together with the band gap for different perturbation strengths. Our results demonstrate that the perturbation-induced geometric distortion of the electronic states breaks the relevant symmetry, giving rise to a 
finite intrinsic NLHC. 


\begin{figure}[h]
	\centering
	\subfigure{\includegraphics[width=0.48\textwidth]{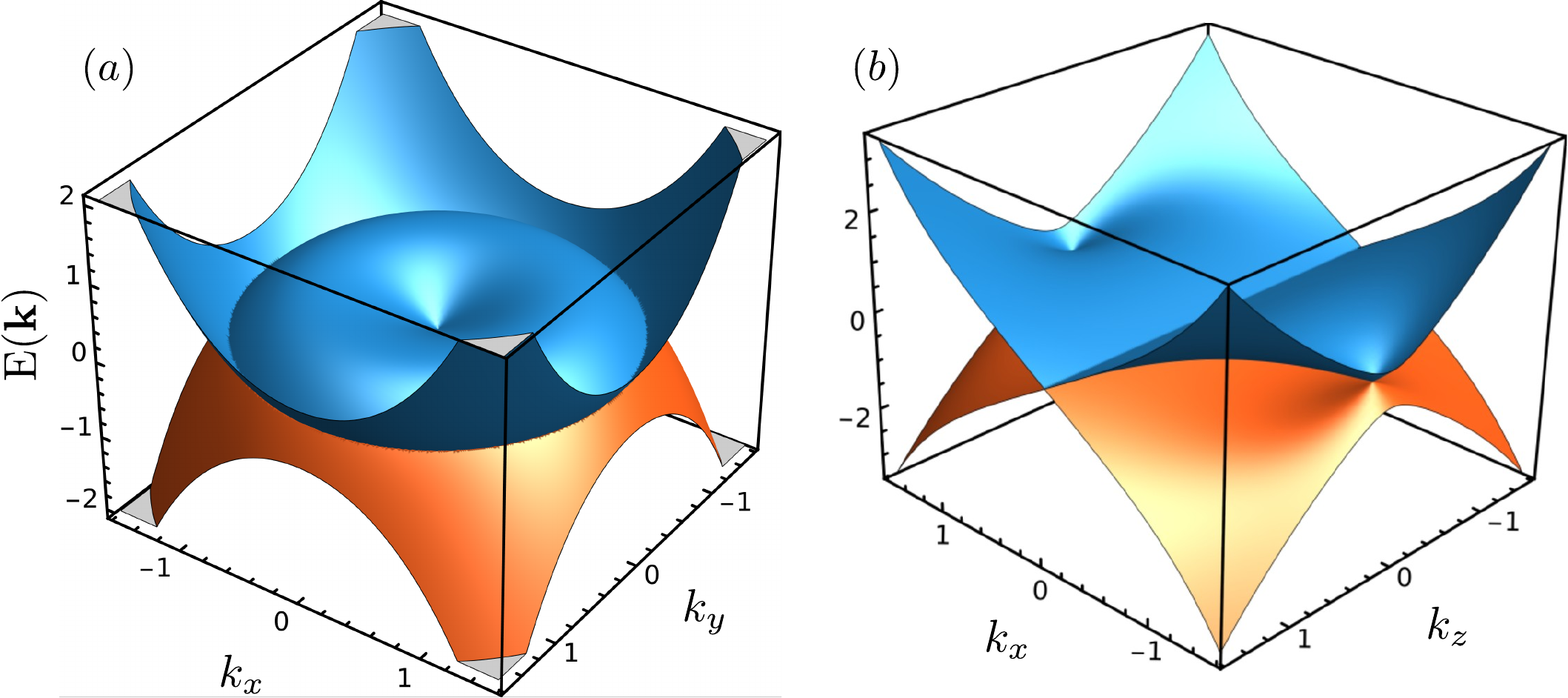}}
	\caption{Electronic band dispersion of the Hopf-link semimetal is demonstrated. (a) Band dispersion is shown in the $k_x$-$k_y$ plane, manifesting a gapless point at the center surrounded by a gapless nodal ring. (b) Band dispersion in the $k_x$-$k_z$ plane is displayed, where a gapless nodal line is visible along the $k_z$ axis, accompanied by two additional gapless points located on either side of the line. The mass parameter is fixed at $m=1$ for both the panels. 
	}
	\label{bands_dispers1}
\end{figure}

\textcolor{blue}{\textit{Low-energy model Hamiltonian for Hopf-link semimetal:}-} 
Here, we present the low-energy continuum Hamiltonian for the Hopf-link semimetal with relevant symmetry properties. This can be written as~\cite{Hopf1,Hopf4}
\begin{eqnarray} 
	H(k) &= [(m - C k^{2}) k_{y} + k_{x} k_{z}] \sigma_{x} \non \\
   &+  [(m - C k^{2}) k_{x} - k_{y} k_{z}] \sigma_{y}\ ,
	\label{Hamiltonian1}
\end{eqnarray}	
where, $\mathbf{k}=(k_x,k_y,k_z)$ is the crystal momentum, $k^{2}=k_x^{2}+k_y^{2}+k_z^{2}$, $C=0.5$, and $\sigma_x$ and $\sigma_y$ denote the Pauli matrices.

The corresponding electronic band structure is shown in Fig.~\ref{bands_dispers1}. In Fig.~\ref{bands_dispers1}(a), we present the band dispersion in the $k_x$--$k_y$ plane, where the conduction and valence bands touch at the Brillouin-zone center as well as along a nodal ring surrounding the center. To illustrate 
the three-dimensional nature of the band crossings, we also display the band dispersion in the $k_x$-$k_z$ plane [see Fig.~\ref{bands_dispers1}(b)]. In this plane, 
the bands remain degenerate along the $k_z$ axis, accompanied by two additional gapless points located symmetrically on either side of the nodal line. 

\begin{figure}[t]
	\centering
	\subfigure{\includegraphics[width=0.5\textwidth]{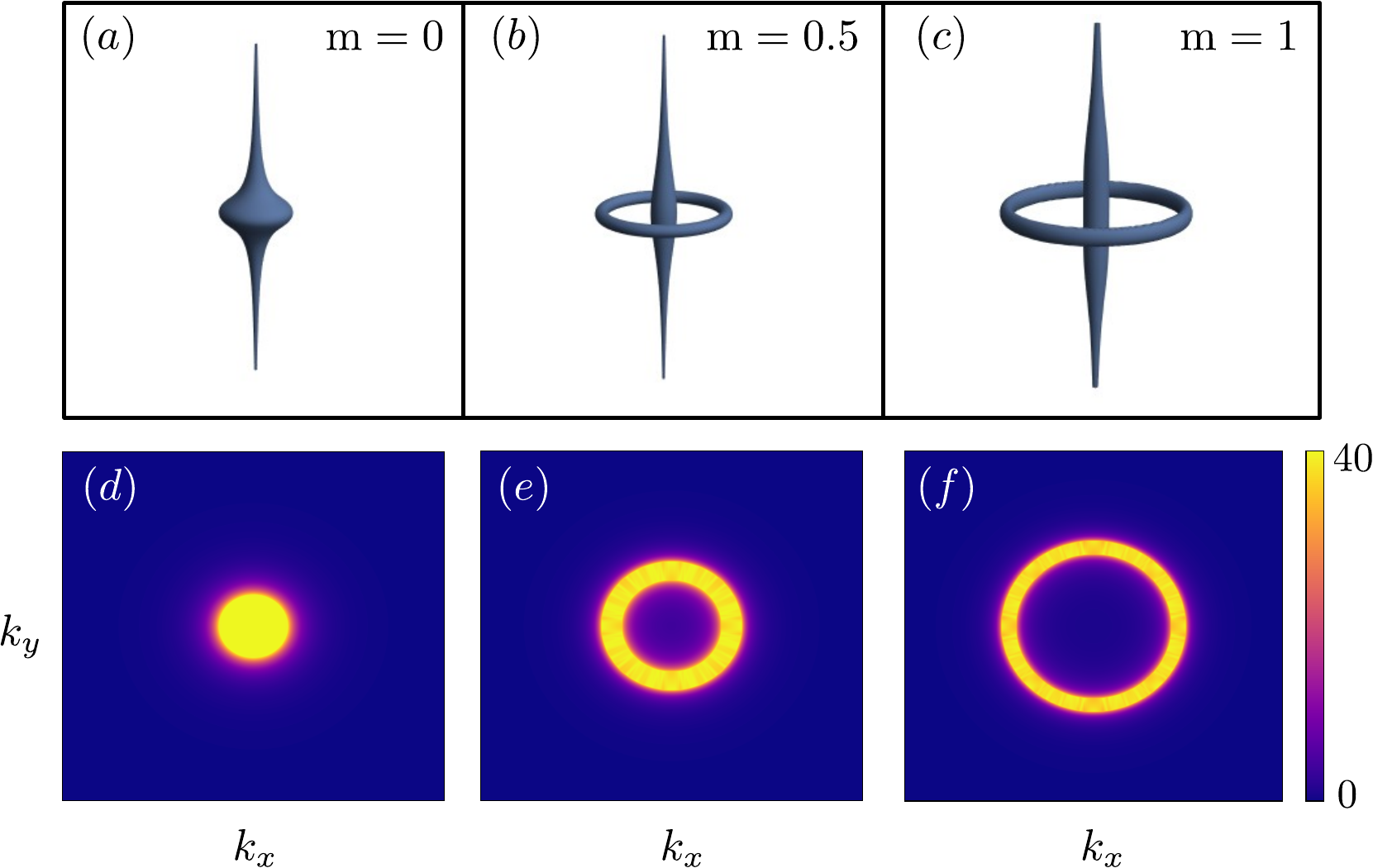}}
	\caption{Link-unlink transition and the corresponding quantum metric in a Hopf-link semimetal are displayed. The upper panels [(a)-(c)] show the evolution 
	of the nodal structure for $m=0$, $0.5$, and $1.0$, respectively. The case $m=0$ corresponds to the unlinked phase, whereas $m>0$ refers to the linked Hopf-link semimetal. The lower panels [(d)-(f)] present the corresponding $zz$-component of the quantum metric, $g_{zz}$, in the $k_x$-$k_y$ plane with $k_z=0$ for the same values of $m$. In the linked phase, $g_{zz}$ develops a ring-shaped structure whose radius increases with the mass parameter, reflecting the evolution of the nodal loop across the link-unlink transition.}
	\label{link_quantum_metric}
\end{figure}

The topology of the nodal structure depends sensitively on the mass parameter $m$. For $m>0$, the system hosts two linked nodal loops: one lies in the $k_z=0$ plane and satisfies $k^{2}=m/C$, while the other is the straight nodal line defined by $k_x=k_y=0$. In Figs.~\ref{link_quantum_metric}(b) and \ref{link_quantum_metric}(c), we illustrate the linked nodal structures for $m=0.5$ and $m=1.0$, respectively. As $m$ is reduced to zero, the system undergoes a topological link-unlink transition, as shown in Fig.~\ref{link_quantum_metric}(a). We also discuss the corresponding lattice models and their connection to the low-energy continuum model 
of the Hopf semimetal in the Supplementary Material (SM)~\cite{supp}.

Finally, we analyze the symmetry properties of our considered model. The bare Hamiltonian preserves a two-fold rotational symmetry ($C_{2z}$), under which $(k_x,k_y,k_z)\xrightarrow{C_{2z}}(-k_x,-k_y,k_z)$
with the corresponding representation ($C_{2z}=\sigma_z$). Accordingly, the Hamiltonian satisfies
\begin{equation}
	C_{2z}H(\mathbf{k})C_{2z}^{-1}
	=H(-k_x,-k_y,k_z)\ .
\end{equation}
The Hamiltonian, however, individually breaks both parity ($\mathcal{P}$) and time-reversal ($\mathcal{T}$) symmetries. In the present two-band representation, 
these operations are represented by, $\mathcal{P}=\sigma_x,\mathcal{T}=K,$
where $K$ denotes complex conjugation. Nevertheless, their combined operation, $\mathcal{PT}=\sigma_xK,$ is preserved. Indeed,
\begin{equation}
	(\sigma_xK)H(\mathbf{k})(\sigma_xK)^{-1}
	=H(\mathbf{k})\ ,
\end{equation}
Since [$(\mathcal{PT})^2=+1$], the preserved ($\mathcal{PT}$) symmetry enforces a vanishing Berry curvature for each nondegenerate band. Consequently, the conventional BCD contribution to the nonlinear Hall response vanishes. The model therefore provides a suitable platform for investigating intrinsic nonlinear responses arising from quantum geometry, particularly those associated with the quantum metric and not captured by the conventional Berry-curvature-dipole mechanism. We discuss these quantum-metric-induced nonlinear effects in the subsequent text.



\textcolor{blue}{\textit{Quantum geometry of Hopf-link semimetal:}-} To begin with, we derive the QGT for a generic two-band Hamiltonian, from which the quantum metric and Berry curvature are obtained as its symmetric and antisymmetric components, respectively. To this end, we consider a generic two-band Hamiltonian of 
the form,
\begin{eqnarray}
	H(\mathbf{k}) &=&  d_{0} (\mathbf{k}) + \mathbf{d} (\mathbf{k}) \cdot \sigma\ ,
\end{eqnarray}
where, $\mathbf{\sigma} = \big( \sigma_{x}, \sigma_{y}, \sigma_{z} \big)$ represents the vector of the Pauli matrix and $\mathbf{d} (\mathbf{k}) = ( d_{x}(\mathbf{k}), d_{y}(\mathbf{k}), d_{z}(\mathbf{k}) )$. With this we define the unit vector, $\mathbf{n} = \frac{\mathbf{d(\mathbf{k})}}{|\mathbf{d}(\mathbf{k})|}$ and express QGT in terms of it as following \cite{QGT_Op_conduct3,QGT_Op_conduct4},
\begin{eqnarray}\label{QuanMetric}
	Q_{\mu \nu}^{\xi} &=& \frac{1}{4} \partial_{\mu} \mathbf{n} \cdot \partial_{\nu} \mathbf{n} - \frac{i}{4} \mathbf{n} \cdot (\partial_{\mu} \mathbf{n} \times \partial_{\nu} \mathbf{n})\ , \non \\
	&=& \text{g}_{\mu \nu}^{\xi} - \frac{i}{2} \Omega_{\mu \nu}^{\xi}\ .
\end{eqnarray}
Here, $\xi$ takes values $\pm 1$ for the two bands $\epsilon_{\pm} = d_{0}(\mathbf{k}) \pm |\mathbf{d}(\mathbf{k})|$. Also, $g_{\mu\nu}^{\xi}$  and $\Omega_{\mu \nu}^{\xi}$ represent the quantum metric and Berry curvature respectively.

As discussed in the previous section, the nodal ring of the Hopf-link semimetal lies in the $k_x$-$k_y$ plane, with its radius tuned by the mass parameter $m$. Consequently, variation $m$ drives the link-unlink transition of the nodal structure. Using the Hamiltonian in Eq.~(\ref{Hamiltonian1}), we obtain the $zz$-component of the quantum metric as~\cite{QGT_Op_conduct3,QGT_Op_conduct4}
\begin{eqnarray}
	g_{zz}=\frac{\left[m-C\left(k_x^2+k_y^2-k_z^2\right)\right]^2}
	{\left[\left(m-C k^2\right)^2+k_z^2\right]^2}\ ,
	\label{gzz}
\end{eqnarray}
where $k^2=k_x^2+k_y^2+k_z^2$.

To visualize its momentum-space profile, we show $g_{zz}$ in the $k_x$-$k_y$ plane by setting $k_z=0$. The corresponding density plots for different values of the mass parameter $m$ are shown in Figs.~\ref{link_quantum_metric}(d)-(f).
For finite values of $m$, $g_{zz}$ develops a pronounced ring-shaped structure, reflecting the presence of the nodal ring in the electronic band structure. The radius of the ring increases as one enhances $m$ (see Figs.~\ref{link_quantum_metric}(e)-(f)). However, as $m$ decreases, the radius of this ring shrinks, mirroring the evolution of the nodal loop across the link-unlink transition. Note that, $m=0$ corresponds to the unlink condition (see Fig.~\ref{link_quantum_metric}(d)). 
This behavior follows directly from Eq.~(\ref{gzz}), where $g_{zz}$ diverges at the band-touching points. In particular, for $k_z=0$, the divergence occurs when $m-Ck^2=0$, corresponding to the nodal-ring condition
\begin{equation}
	k_x^2+k_y^2=\frac{m}{C}\ ,
\end{equation}
Thus, the momentum-space distribution of the quantum metric faithfully captures the geometry and evolution of the nodal structure of the band. For completeness, 
we present the corresponding distributions of other components of quantum metrics $g_{xx}$, $g_{xy}$, $g_{zz}$, $g_{zx}$ in the $k_x$-$k_y$ and $k_x$-$k_z$ 
planes in SM~\cite{supp}.

\textcolor{blue}{\textit{Optical conductivity:}-} In this part, we express the optical conductivity in terms of the quantum metric and investigate its evolution as the model parameters are tuned across the Hopf link-unlink transition.

For a two-band system, the linear optical conductivity in terms of the QGT is given by~\cite{QGT_Op_conduct3,QGT_Op_conduct4},
\begin{equation}
	\sigma_{\mu \nu} = \pi \omega e^{2} \int \frac{d^{3}k}{(2\pi)^{3}} \Delta f  \delta (\hbar \omega - \Delta \varepsilon) Q_{\mu \nu}^{\pm}\ ,
\end{equation}
where, $\Delta \varepsilon = (\varepsilon_{+}(\mathbf{k}) - \varepsilon_{-}(\mathbf{k}))$ denotes the energy difference between the two bands (\ie band gap), and $\Delta f = f_{-} - f_{+}$ is the difference between the Fermi-Dirac distribution functions of the corresponding bands. Finally, we obtain the real part of the optical conductivity in terms of the quantum metric~\cite{QGT_Op_conduct3},
\begin{equation}
	\text{Re}[\sigma_{\mu \nu}] = \pi \omega e^{2} \int \frac{d^{3}k}{(2\pi)^{3}} \Delta f  \delta (\hbar \omega - \Delta \varepsilon) g_{\mu \nu}^{\pm}\ ,
	\label{opt.sigma}
\end{equation}
From Eq.~(\ref{opt.sigma}) we note that, when the photon energy ($\hbar \omega$) is smaller than the band gap ($\Delta \varepsilon$) \ie $\hbar \omega < \Delta \varepsilon$, the optical absorption is zero. The frequency at which $\omega_{0} = \frac{\Delta \varepsilon}{\hbar}$ is coined as the band edge frequency.

\begin{figure*}[t]
	\subfigure{\includegraphics[width=0.9\textwidth]{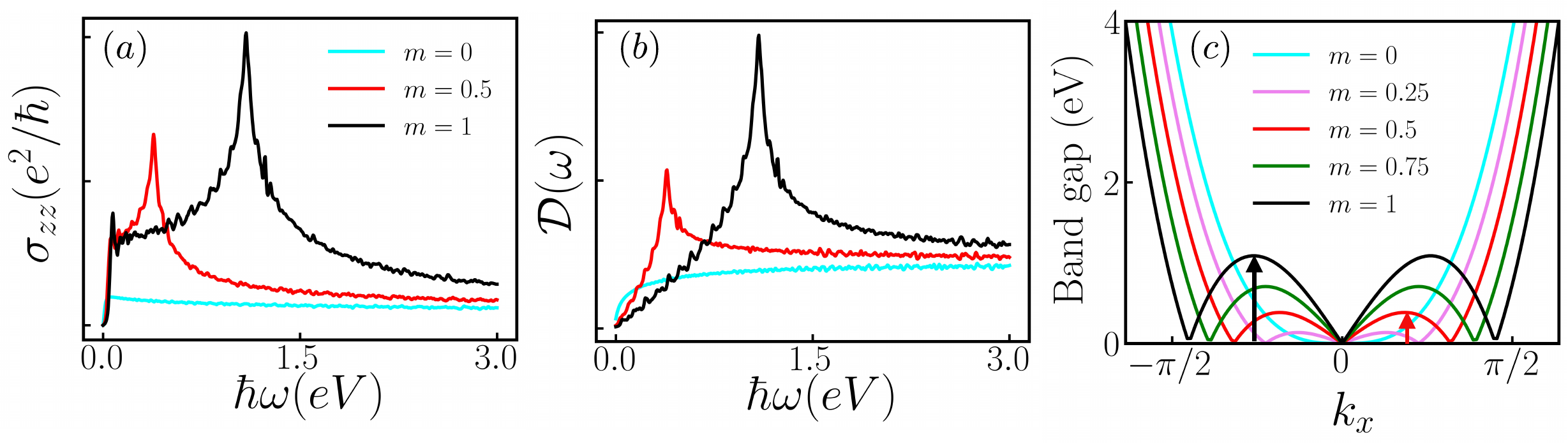}}
	\caption{(a) Interband optical conductivity along the $z$ direction, $\sigma_{zz}$, is displayed as a function of the photon frequency for $m=0$, $0.5$, and $1.0$, indicated by the cyan, red, and black curves, respectively. Pronounced peaks appear for finite values of the mass parameters. (b) Interband JDOS for the same values of $m$ is shown using the same color schemes. The JDOS peaks coincide with those in the optical conductivity, confirming their common origin. (c) Band gap, $\Delta\varepsilon$, is depicted as a function of $k_x$ with $k_y=k_z=0$ for different values of the mass parameter $m$. The curves correspond to $m=0$, $0.25$, $0.5$, $0.75$, and $1.0$, are shown in cyan, purple, red, green, and black colors, respectively. Three gapless points are observed: one at $k_x=0$, corresponding to the nodal line, and two at finite $k_x$, corresponding to the nodal ring. The arrows indicate the maximum band gap between the nodal line and the nodal ring, which precisely determines the frequency at which the peaks in both the optical conductivity and the JDOS occur.}
	\label{OPC_JDOS_BG}
\end{figure*}
As discussed in the earlier text, 
the nodal loop of the Hopf-link semimetal lies in the $k_x$-$k_y$ plane. Therefore, we focus on the $zz$-component of the quantum metric and, correspondingly, on the real part of the optical conductivity along the $z$-direction. In Fig.~\ref{OPC_JDOS_BG}(a), we showcase the $zz$-component of the optical conductivity, $\sigma_{zz}$, as a function of the photon frequency $\omega$ for three different values of the Hopf mass term ($m$). For finite values of $m$, the optical conductivity exhibits pronounced sharp peaks, whose positions shift to higher frequencies with increasing $m$. In contrast, no such peak is observed for $m=0$, corresponding to the unlinked phase. Thus, the presence (absence) of sharp peaks in the optical conductivity provides a clear signature of the linked (unlinked) phase of the Hopf-link semimetal.

To understand the origin of these peaks, we then calculate the interband JDOS and analyze the band-gap profile. The corresponding JDOS is shown in Fig.~\ref{OPC_JDOS_BG}(b) for the same set of mass parameters, using the same set of color scheme as depicted in Fig.~\ref{OPC_JDOS_BG}(a). We find that the JDOS peaks occur at exactly the same frequencies as the peaks in the optical conductivity, establishing a direct correspondence between the two quantities.

Further insight can be gained by examining the momentum-dependent band gap. In Fig.~\ref{OPC_JDOS_BG}(c), we show the band gap, $\Delta\varepsilon$, as a function of $k_x$ by fixing $k_y=k_z=0$ and considering different values of $m$. The underlying Hopf-link nodal structure is clearly reflected in the band-gap profile. The gap vanishes at $k_x=0$, corresponding to the nodal line, whereas for $m>0$ it also closes at finite values of $k_x$, signaling the presence of the nodal loop. The maximum band gap between the nodal line and the nodal loop, indicated by the arrows in Fig.~\ref{OPC_JDOS_BG}(c), precisely determines the frequencies at which the peaks in both the optical conductivity and the JDOS appear. Note that, the maximum band gap increases as one enhances the mass parameter $m$ resulting in larger
peak height in the optical conductivity and JDOS. This demonstrates that the characteristic optical response of the Hopf-link semimetal is directly governed by its underlying band structure.

\textcolor{blue}{\textit{Quantum metric dipole induced non-linear response:}-} Here, we demonstrate the emergence of a finite QMD-induced nonlinear Hall response upon introducing a perturbative term into the Hamiltonian.

\begin{figure}[t]
	\centering
	\subfigure{\includegraphics[width=0.48\textwidth]{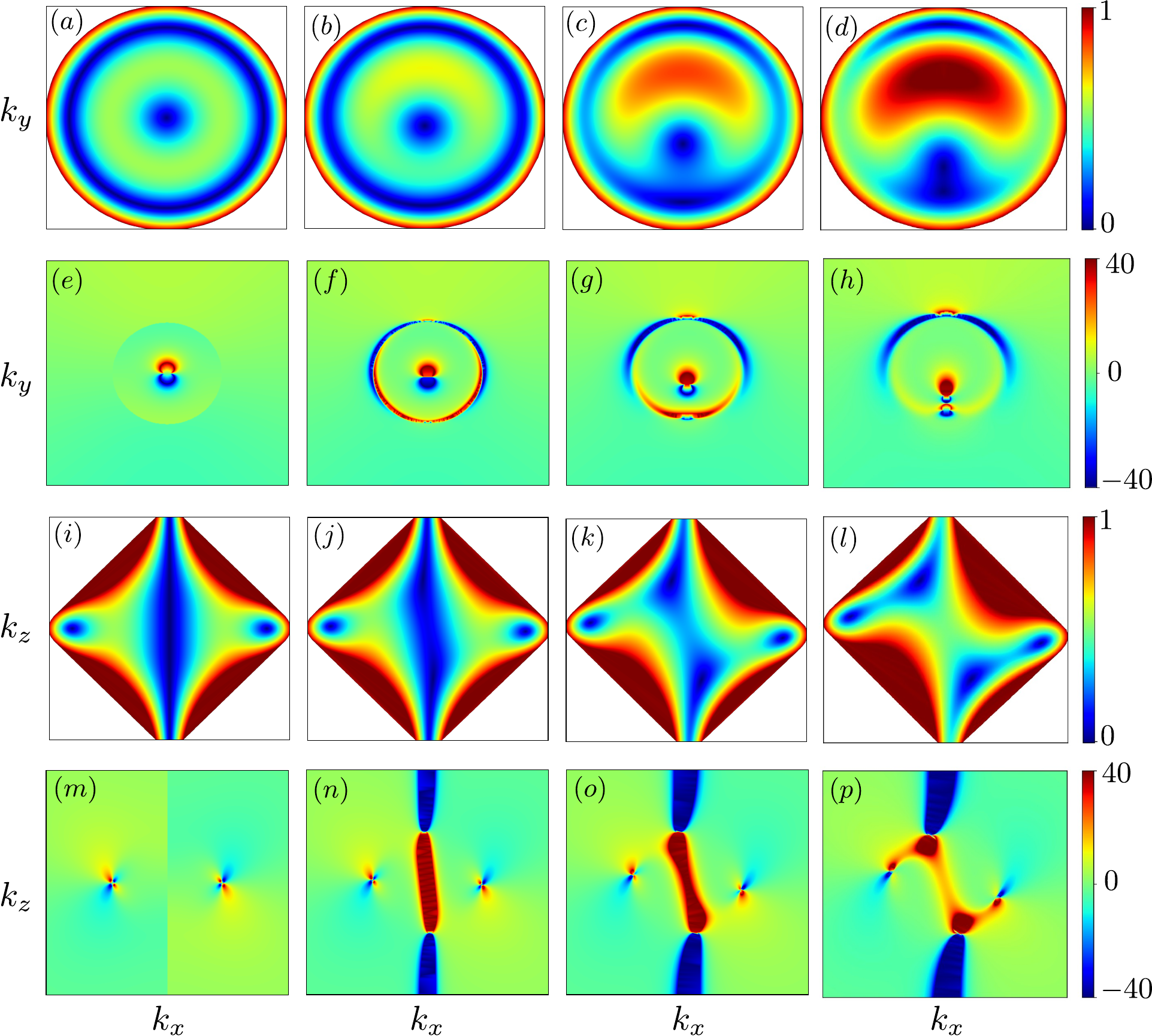}}
	\caption{Momentum-space distributions of the band gap and the QMD kernel, $(v_{y}g_{xx}-v_{x}g_{yx})$, are demonstrated for different values of the perturbation strength $\beta$. Panels (a)-(d) exhibit the band-gap distribution in the $k_x$-$k_y$ plane for $\beta=0$, $0.1$, $0.3$, and $0.5$, respectively. The corresponding distributions of the QMD kernel are presented in panels (e)-(h). Panels (i)-(l) display the band-gap distribution in the $k_x$-$k_z$ plane for the same values of $\beta$, while panels (m)-(p) showcase the corresponding momentum-space distributions of the QMD kernel. The progressive distortion of both the nodal structure and the QMD kernel with increasing $\beta$ gives rise to a finite intrinsic nonlinear Hall response (see latter text for discussion).}
	\label{Gap_and_QMD}
\end{figure}

The quantum metric dipole associated with the $n^{\mathrm{th}}$ band can be defined as~\cite{NLHC_In4}
\begin{equation}
	D^{n}_{QM} = \int \frac{d^{3}k}{(2\pi)^{3}} (v_{y}^{n} g^{n}_{xx} - v_{x}^{n} g^{n}_{yx})\delta(E_{n}-\mu)\ ,
\end{equation}
where, $E_{n}$ is the energy of the $n^{\text{th}}$ band and $\mu$ represents the chemical potential. Here,
$v_{x,y}^{n}$ denote the group velocities of electrons along $x$ and $y$ directions in the $n^{\text{th}}$ band. 
Also, $g^{n}_{xx}$ and $g^{n}_{yx}$ are the quantum metrics and can be found from Eq.~(\ref{QuanMetric}). We depict the behavior of this integral kernel 
in Fig.~\ref{Gap_and_QMD} choosing different momentum space planes. 

By itself, the Hamiltonian in Eq.~(\ref{Hamiltonian1}) yields a vanishing intrinsic nonlinear Hall response. A nonzero nonlinear Hall conductivity can be engineered by introducing a symmetry-breaking perturbation of the form $H'=\beta\sigma_x$, where $\beta$ denotes the perturbation strength. To understand its impact, 
in Fig.~\ref{Gap_and_QMD}, we show the evolution of the band gap together with the momentum-space distribution of the QMD kernel for different values of $\beta$. 
In Figs.~\ref{Gap_and_QMD}(a)-(d), we illustrate the band-gap distribution in the $k_x$-$k_y$ plane for increasing values of $\beta$. In the absence of the perturbation ($\beta=0$), the appearance of dark-blue ring corresponds to the nodal loop of the Hopf link, while the gapless point at the center originates from the nodal line extending along the $k_z$ axis. As the perturbation strength increases, the nodal structure is progressively distorted, although it remains gapless over the parameter range considered. In the next row (\ie in Figs.~\ref{Gap_and_QMD}(e)-(h)), the momentum-space distribution of the QMD kernel,
$(v_y^n g_{xx}^n-v_x^n g_{yx}^n)$, is shown for the same respective values of $\beta$. As discussed earlier, the bare model (\ie $\beta = 0$) possesses the $C_{2z}$ symmetry, that maps $(k_x,k_y,k_z) \to (-k_x,-k_y,k_z)$. Consequently, we find that the momentum-space distribution of the QMD kernel, $(v_y^n g_{xx}^n - v_x^n g_{yx}^n)$, is odd under this transformation. This antisymmetric behavior is also evident in Fig.~\ref{Gap_and_QMD}(e): for any momentum point $(k_x,k_y)$, the QMD kernel at the corresponding point $(-k_x,-k_y)$ takes the opposite value. As $\beta$ increases, this symmetry is gradually broken, producing an increasingly asymmetric distribution of the QMD kernel. To further illustrate this behavior, we examine the same quantities in the $k_x$-$k_z$ plane by fixing $k_y=0$. The corresponding band-gap distributions are shown in Figs.~\ref{Gap_and_QMD}(i)-(l). For $\beta=0$, the gapless line along the $k_z$ axis represents the nodal line of the Hopf link, while the two gapless points located symmetrically on either side correspond to the intersection of the nodal loop with the $k_x$-$k_z$ plane. Similar to the $k_x$-$k_y$ plane, the nodal structure undergoes a gradual deformation with increasing $\beta$. Finally, in Figs.~\ref{Gap_and_QMD}(m)-(p) we display the corresponding QMD kernel in the $k_x$-$k_z$ plane. In the absence of the perturbation, the positive and negative contributions are symmetrically distributed about the $k_x=0$ axis, yielding a zero net QMD. As $\beta$ becomes finite, a pronounced contribution develops along the $k_x=0$ line, and the overall distribution becomes increasingly asymmetric. This perturbation-induced distortion of the quantum metric dipole ultimately gives rise to a finite intrinsic nonlinear Hall conductivity.

\begin{figure}[b]
	\subfigure{\includegraphics[width=0.48\textwidth]{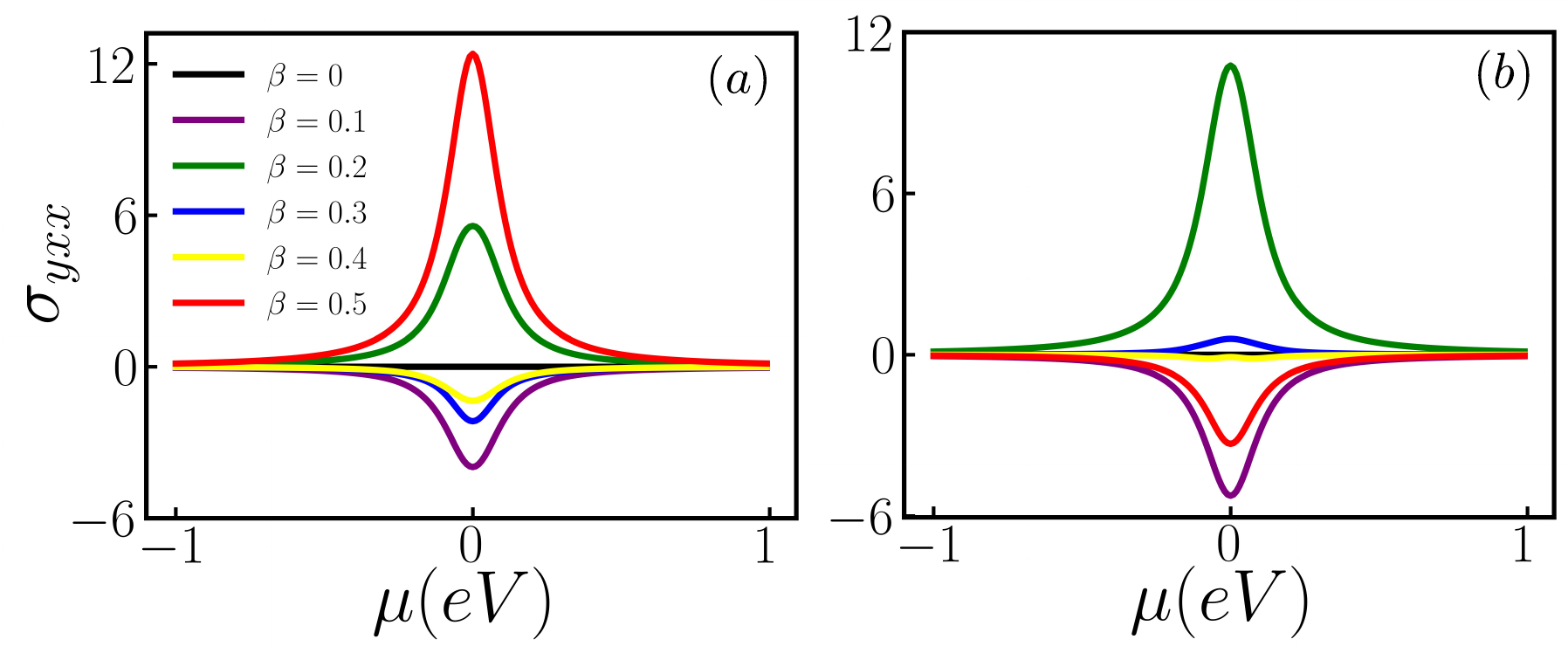}}
	\caption{In panels (a) and (b), we present the QMD-induced nonlinear Hall conductivity, $\sigma_{yxx}$, (in units of $e^{3}/\hbar\cdot\mathrm{eV}$), as a function of the chemical potential $\mu$ (in units of eV) for two different values of the Hopf mass, $m=0.5$ and $m=1$, respectively, considering different values of the perturbation strength $\beta$. The black, violet, green, blue, yellow, and red curves correspond to $\beta=0$, $0.1$, $0.2$, $0.3$, $0.4$, and $0.5$, respectively.
	}
	\label{NL_Hall_response}
\end{figure}

The quantum metric dipole induced non-linear Hall conductivity is defined as~\cite{NLHC_In4,NLHC_In5,NLHC_QMD}, 
\begin{equation}
	\sigma_{yxx} =  -2 e^{3} \sum_{n} \int \frac{d^{3}k}{(2\pi)^{3}} \frac{(v_{y}^{n} g^{n}_{xx} - v_{x}^{n} g^{n}_{yx})}{E_{n} - E_{\bar{n}}} \delta (E_{n} - \mu)\ ,
\end{equation}
As defined earlier, $v_{x,y}^{n}$ denote the group velocity for the $n^{\text{th}}$ band and $g^{n}_{xx}$, $g^{n}_{yx}$ correspond to the quantum metrics. 
Here, $E_{n}$ and $E_{\bar{n}}$ symbolically denote the energy dispersion of the $n^{\rm{th}}$ Bloch band and its complementary band ($n \neq \bar{n}$). For the two-band model considered here, $n = \pm$ labels the conduction and valence bands, respectively, with $\bar{n} = \mp$.

In Fig.~\ref{NL_Hall_response}, we present the QMD-induced nonlinear Hall conductivity as a function of the chemical potential ($\mu$) for different values of the perturbation strength $\beta$. As discussed above, the QMD kernel is odd under the $C_{2z}$ transformation, whereas both the band energy and the delta function are even. Consequently, the contributions arising from $C_{2z}$-related momentum points exactly cancel each other upon integration over momentum space, resulting in a vanishing nonlinear Hall response for the bare model ($\beta=0$). We here focus on the linked phase of the Hopf-link semimetal by considering two representative values of the mass parameter, $m=0.5$ and $m=1.0$, whose corresponding results are shown in Figs.~\ref{NL_Hall_response}(a) and \ref{NL_Hall_response}(b), respectively. In contrast, the unlinked phase does not exhibit any finite nonlinear Hall response. In both panels, the nonlinear Hall conductivity $\sigma_{yxx}$ 
is depicted for different values of perturbation strengths $\beta$, as represented by different colors. For all finite values of $\beta$, $\sigma_{yxx}$ exhibits a pronounced peak near the charge-neutrality point ($\mu=0$). Although the magnitude of peak height of the nonlinear Hall conductivity does not vary monotonically 
with the perturbation strength, the response becomes finite once the perturbation is switched on and may even change sign as $\beta$ is varied. As discussed in the previous section, the perturbation reconstructs the electronic band structure and modifies the momentum-space distribution of the quantum metric dipole. This redistribution of the QMD kernel produces a finite intrinsic nonlinear Hall conductivity in the linked phase of the Hopf-link semimetal.


\textcolor{blue}{\textit{Summary and conclusions:}-} To summarize, we investigate the optical and nonlinear transport properties of a low-energy continuum model 
describing a Hopf-link semimetal from the perspective of quantum geometry. We begin by analyzing the electronic band structure to demonstrate the evolution of the nodal-line topology through the link-to-unlink transition. From the analytical expression of the $zz$-component of the quantum metric, we show that the quantum geometry carries distinct signatures across the topological transition, reflecting the underlying modification in the nodal-line configuration. We then express the interband optical conductivity in terms of quantum metric and track its evolution throughout the link-unlink transition. Our findings reveal that the linked phase exhibits pronounced, sharp peaks in the optical conductivity, which disappear in the unlinked phase. To uncover the microscopic origin of these peaks, we compute the JDOS and identify an excellent correspondence between the JDOS maxima and the optical conductivity peaks. Furthermore, by evaluating the momentum-resolved band gap along representative one-dimensional momentum cuts, we establish that the band-gap structure directly affects both the energy positions and magnitudes of the conductivity peaks. These results demonstrate that the optical response serves as a clear fingerprint of the topological evolution of the nodal lines.

The second part of our work focuses on the QMD and its explicit role in generating the NLHC. Our analysis indicates that the unperturbed Hopf-link semimetal does not exhibit any finite NLHC because of the underlying symmetry constraints. However, introducing an appropriate perturbative term activates a finite QMD, leading to a measurable nonlinear Hall response. By examining the momentum-space distribution of the QMD kernel together with the corresponding band-gap evolution for different perturbation strengths, we find that the geometric distortion of the electronic bands is directly responsible for the emergence and enhancement of the NLHC. Overall, our findings establish a direct connection of the quantum metric properties with optical conductivity and nonlinear Hall transport in Hopf-link semimetals.
 
We demonstrate that the QMD acts as a key driving mechanism for the scattering-independent nonlinear Hall effect. This phenomenon has previously been observed in several materials, including pristine MnBi$_2$Te$_4$ and MnBi$_2$Te$_4$/black phosphorus heterostructures which are topological antiferromagnet~\cite{NLHC_In4,NHLC_expt}. We expect that our findings will stimulate further theoretical and experimental explorations of quantum-geometric phenomena in topological semimetals and related quantum systems.

\textcolor{blue}{\textit{Acknowledgments:}-} K.B. and A.S. acknowledge the two workstations provided by the Institute of Physics, Bhubaneswar from the DAE APEX project for numerical computations. D.C. acknowledges financial support from DST (project number DST/WISE-PDF/PM-40/2023).

\textcolor{blue}{\textit{Data availibility statement:}-} The datasets generated and analyzed during the current study are available from the corresponding author upon reasonable request.


\bibliography{bibfile}{}

@Article{QGT_Provost,
author={Provost, J. P.
and Vallee, G.},
title={Riemannian structure on manifolds of quantum states},
journal={Communications in Mathematical Physics},
year={1980},
month={Sep},
day={01},
volume={76},
number={3},
pages={289-301},
issn={1432-0916},
doi={10.1007/BF02193559},
url={https://doi.org/10.1007/BF02193559}
}

@article{QGT_Paivi,
  title = {Essay: Where Can Quantum Geometry Lead Us?},
  author = {T\"orm\"a, P\"aivi},
  journal = {Phys. Rev. Lett.},
  volume = {131},
  issue = {24},
  pages = {240001},
  numpages = {7},
  year = {2023},
  month = {Dec},
  publisher = {American Physical Society},
  doi = {10.1103/PhysRevLett.131.240001},
  url = {https://link.aps.org/doi/10.1103/PhysRevLett.131.240001}
}

@article{Berry_Phase,
    author = {Berry, Michael Victor},
    title = {Quantal phase factors accompanying adiabatic changes},
    journal = {Proceedings of the Royal Society of London. A. Mathematical and Physical Sciences},
    volume = {392},
    number = {1802},
    pages = {45-57},
    year = {1984},
    month = {03},
    issn = {0080-4630},
    doi = {10.1098/rspa.1984.0023},
    url = {https://doi.org/10.1098/rspa.1984.0023}
}

@article{Berry_Phase_RMP,
  title = {Berry phase effects on electronic properties},
  author = {Xiao, Di and Chang, Ming-Che and Niu, Qian},
  journal = {Rev. Mod. Phys.},
  volume = {82},
  issue = {3},
  pages = {1959--2007},
  numpages = {0},
  year = {2010},
  month = {Jul},
  publisher = {American Physical Society},
  doi = {10.1103/RevModPhys.82.1959},
  url = {https://link.aps.org/doi/10.1103/RevModPhys.82.1959}
}

@article{Thouless,
  title = {Quantization of particle transport},
  author = {Thouless, D. J.},
  journal = {Phys. Rev. B},
  volume = {27},
  issue = {10},
  pages = {6083--6087},
  numpages = {0},
  year = {1983},
  month = {May},
  publisher = {American Physical Society},
  doi = {10.1103/PhysRevB.27.6083},
  url = {https://link.aps.org/doi/10.1103/PhysRevB.27.6083}
}

@article{Resta,
doi = {10.1088/0953-8984/22/12/123201},
url = {https://doi.org/10.1088/0953-8984/22/12/123201},
year = {2010},
month = {mar},
publisher = {},
volume = {22},
number = {12},
pages = {123201},
author = {Resta, Raffaele},
title = {Electrical polarization and orbital magnetization: the modern theories},
journal = {Journal of Physics: Condensed Matter}
}

@article{Ob_mag1,
  title = {Geometrical effects in orbital magnetic susceptibility},
  author = {Gao, Yang and Yang, Shengyuan A. and Niu, Qian},
  journal = {Phys. Rev. B},
  volume = {91},
  issue = {21},
  pages = {214405},
  numpages = {12},
  year = {2015},
  month = {Jun},
  publisher = {American Physical Society},
  doi = {10.1103/PhysRevB.91.214405},
  url = {https://link.aps.org/doi/10.1103/PhysRevB.91.214405}
}

@article{Ob_mag2,
  title = {Strain Engineering of the Berry Curvature Dipole and Valley Magnetization in Monolayer ${\mathrm{MoS}}_{2}$},
  author = {Son, Joolee and Kim, Kyung-Han and Ahn, Y. H. and Lee, Hyun-Woo and Lee, Jieun},
  journal = {Phys. Rev. Lett.},
  volume = {123},
  issue = {3},
  pages = {036806},
  numpages = {6},
  year = {2019},
  month = {Jul},
  publisher = {American Physical Society},
  doi = {10.1103/PhysRevLett.123.036806},
  url = {https://link.aps.org/doi/10.1103/PhysRevLett.123.036806}
}

@article{Hall_effects1,
  title = {Anomalous Hall effect},
  author = {Nagaosa, Naoto and Sinova, Jairo and Onoda, Shigeki and MacDonald, A. H. and Ong, N. P.},
  journal = {Rev. Mod. Phys.},
  volume = {82},
  issue = {2},
  pages = {1539--1592},
  numpages = {0},
  year = {2010},
  month = {May},
  publisher = {American Physical Society},
  doi = {10.1103/RevModPhys.82.1539},
  url = {https://link.aps.org/doi/10.1103/RevModPhys.82.1539}
}

@article{Hall_effects2,
  title = {Quantum Nonlinear Hall Effect Induced by Berry Curvature Dipole in Time-Reversal Invariant Materials},
  author = {Sodemann, Inti and Fu, Liang},
  journal = {Phys. Rev. Lett.},
  volume = {115},
  issue = {21},
  pages = {216806},
  numpages = {5},
  year = {2015},
  month = {Nov},
  publisher = {American Physical Society},
  doi = {10.1103/PhysRevLett.115.216806},
  url = {https://link.aps.org/doi/10.1103/PhysRevLett.115.216806}
}

@Article{Hall_effects3,
author={Kang, Kaifei
and Li, Tingxin
and Sohn, Egon
and Shan, Jie
and Mak, Kin Fai},
title={Nonlinear anomalous Hall effect in few-layer WTe2},
journal={Nature Materials},
year={2019},
month={Apr},
day={01},
volume={18},
number={4},
pages={324-328},
issn={1476-4660},
doi={10.1038/s41563-019-0294-7},
url={https://doi.org/10.1038/s41563-019-0294-7}
}

@article{Hall_effects4,
doi={10.1126/science.1250140},
author = {K. F. Mak  and K. L. McGill  and J. Park  and P. L. McEuen },
title = {The valley Hall effect in MoS<sub>2</sub> transistors},
journal = {Science},
volume = {344},
number = {6191},
pages = {1489-1492},
year = {2014},
doi = {10.1126/science.1250140},
URL = {https://www.science.org/doi/abs/10.1126/science.1250140},
}

@Article{Hall_effects5,
author={Lee, Jieun
and Mak, Kin Fai
and Shan, Jie},
title={Electrical control of the valley Hall effect in bilayer MoS2 transistors},
journal={Nature Nanotechnology},
year={2016},
month={May},
day={01},
volume={11},
number={5},
pages={421-425},
issn={1748-3395},
doi={10.1038/nnano.2015.337},
url={https://doi.org/10.1038/nnano.2015.337}
}

@article{Hall_effects6,
  title = {Spin Hall effects},
  author = {Sinova, Jairo and Valenzuela, Sergio O. and Wunderlich, J. and Back, C. H. and Jungwirth, T.},
  journal = {Rev. Mod. Phys.},
  volume = {87},
  issue = {4},
  pages = {1213--1260},
  numpages = {47},
  year = {2015},
  month = {Oct},
  publisher = {American Physical Society},
  doi = {10.1103/RevModPhys.87.1213},
  url = {https://link.aps.org/doi/10.1103/RevModPhys.87.1213}
}

@Article{SF_SC_flat_band1,
author={Peotta, Sebastiano
and T{\"o}rm{\"a}, P{\"a}ivi},
title={Superfluidity in topologically nontrivial flat bands},
journal={Nature Communications},
year={2015},
month={Nov},
day={20},
volume={6},
number={1},
pages={8944},
issn={2041-1723},
doi={10.1038/ncomms9944},
url={https://doi.org/10.1038/ncomms9944}
}

@Article{SF_SC_flat_band2,
author={T{\"o}rm{\"a}, P{\"a}ivi
and Peotta, Sebastiano
and Bernevig, Bogdan A.},
title={Superconductivity, superfluidity and quantum geometry in twisted multilayer systems},
journal={Nature Reviews Physics},
year={2022},
month={Aug},
day={01},
volume={4},
number={8},
pages={528-542},
issn={2522-5820},
doi={10.1038/s42254-022-00466-y},
url={https://doi.org/10.1038/s42254-022-00466-y}
}

@Article{SF_SC_flat_band3,
author={Tian, Haidong
and Gao, Xueshi
and Zhang, Yuxin
and Che, Shi
and Xu, Tianyi
and Cheung, Patrick
and Watanabe, Kenji
and Taniguchi, Takashi
and Randeria, Mohit
and Zhang, Fan
and Lau, Chun Ning
and Bockrath, Marc W.},
title={Evidence for Dirac flat band superconductivity enabled by quantum geometry},
journal={Nature},
year={2023},
month={Feb},
day={01},
volume={614},
number={7948},
pages={440-444},
issn={1476-4687},
doi={10.1038/s41586-022-05576-2},
url={https://doi.org/10.1038/s41586-022-05576-2}
}

@article{SF_SC_flat_band4,
  title = {Ginzburg-Landau Theory of Flat-Band Superconductors with Quantum Metric},
  author = {Chen, Shuai A. and Law, K. T.},
  journal = {Phys. Rev. Lett.},
  volume = {132},
  issue = {2},
  pages = {026002},
  numpages = {7},
  year = {2024},
  month = {Jan},
  publisher = {American Physical Society},
  doi = {10.1103/PhysRevLett.132.026002},
  url = {https://link.aps.org/doi/10.1103/PhysRevLett.132.026002}
}

@article{QPT,
  title = {Fidelity susceptibility and long-range correlation in the Kitaev honeycomb model},
  author = {Yang, Shuo and Gu, Shi-Jian and Sun, Chang-Pu and Lin, Hai-Qing},
  journal = {Phys. Rev. A},
  volume = {78},
  issue = {1},
  pages = {012304},
  numpages = {6},
  year = {2008},
  month = {Jul},
  publisher = {American Physical Society},
  doi = {10.1103/PhysRevA.78.012304},
  url = {https://link.aps.org/doi/10.1103/PhysRevA.78.012304}
}

@article{Bound_topological_gap,
  title = {Fundamental Bound on Topological Gap},
  author = {Onishi, Yugo and Fu, Liang},
  journal = {Phys. Rev. X},
  volume = {14},
  issue = {1},
  pages = {011052},
  numpages = {12},
  year = {2024},
  month = {Mar},
  publisher = {American Physical Society},
  doi = {10.1103/PhysRevX.14.011052},
  url = {https://link.aps.org/doi/10.1103/PhysRevX.14.011052}
}

@article{Noise,
  title = {Measuring the quantum geometry of Bloch bands with current noise},
  author = {Neupert, Titus and Chamon, Claudio and Mudry, Christopher},
  journal = {Phys. Rev. B},
  volume = {87},
  issue = {24},
  pages = {245103},
  numpages = {5},
  year = {2013},
  month = {Jun},
  publisher = {American Physical Society},
  doi = {10.1103/PhysRevB.87.245103},
  url = {https://link.aps.org/doi/10.1103/PhysRevB.87.245103}
}

@article{QGT_Op_conduct1,
  title = {Fundamental Bound on Topological Gap},
  author = {Onishi, Yugo and Fu, Liang},
  journal = {Phys. Rev. X},
  volume = {14},
  issue = {1},
  pages = {011052},
  numpages = {12},
  year = {2024},
  month = {Mar},
  publisher = {American Physical Society},
  doi = {10.1103/PhysRevX.14.011052},
  url = {https://link.aps.org/doi/10.1103/PhysRevX.14.011052}
}

@article{QGT_Op_conduct2,
    author = {Berry, Michael Victor},
    title = {Quantal phase factors accompanying adiabatic changes},
    journal = {Proceedings of the Royal Society of London. A. Mathematical and Physical Sciences},
    volume = {392},
    number = {1802},
    pages = {45-57},
    year = {1984},
    month = {03},
    issn = {0080-4630},
    doi = {10.1098/rspa.1984.0023},
    url = {https://doi.org/10.1098/rspa.1984.0023}
}

@article{QGT_Op_conduct3,
  title = {Analytic approach to quantum metric and optical conductivity in Dirac models with parabolic mass in arbitrary dimensions},
  author = {Ezawa, Motohiko},
  journal = {Phys. Rev. B},
  volume = {110},
  issue = {19},
  pages = {195437},
  numpages = {11},
  year = {2024},
  month = {Nov},
  publisher = {American Physical Society},
  doi = {10.1103/PhysRevB.110.195437},
  url = {https://link.aps.org/doi/10.1103/PhysRevB.110.195437}
}

@article{QGT_Op_conduct4,
  title = {Quantum geometry and low-frequency optical conductivity of nodal planes},
  author = {Wiedmann, Raymond and Alpin, Kirill and Hirschmann, Moritz M. and Schnyder, Andreas P.},
  journal = {Phys. Rev. B},
  volume = {112},
  issue = {12},
  pages = {125102},
  numpages = {11},
  year = {2025},
  month = {Sep},
  publisher = {American Physical Society},
  doi = {10.1103/2f4l-tz7p},
  url = {https://link.aps.org/doi/10.1103/2f4l-tz7p}
}

@article{Op_conduc_Dirac_Weyl1,
  title = {Optical spectroscopy study of the three-dimensional Dirac semimetal ${\mathrm{ZrTe}}_{5}$},
  author = {Chen, R. Y. and Zhang, S. J. and Schneeloch, J. A. and Zhang, C. and Li, Q. and Gu, G. D. and Wang, N. L.},
  journal = {Phys. Rev. B},
  volume = {92},
  issue = {7},
  pages = {075107},
  numpages = {5},
  year = {2015},
  month = {Aug},
  publisher = {American Physical Society},
  doi = {10.1103/PhysRevB.92.075107},
  url = {https://link.aps.org/doi/10.1103/PhysRevB.92.075107}
}

@article{Op_conduc_Dirac_Weyl2,
  title = {Interband optical conductivity of the [001]-oriented Dirac semimetal ${\mathrm{Cd}}_{3}{\mathrm{As}}_{2}$},
  author = {Neubauer, D. and Carbotte, J. P. and Nateprov, A. A. and L\"ohle, A. and Dressel, M. and Pronin, A. V.},
  journal = {Phys. Rev. B},
  volume = {93},
  issue = {12},
  pages = {121202(R)},
  numpages = {5},
  year = {2016},
  month = {Mar},
  publisher = {American Physical Society},
  doi = {10.1103/PhysRevB.93.121202},
  url = {https://link.aps.org/doi/10.1103/PhysRevB.93.121202}
}

@article{Op_conduc_Dirac_Weyl3,
  title = {Optical spectroscopy of the Weyl semimetal TaAs},
  author = {Xu, B. and Dai, Y. M. and Zhao, L. X. and Wang, K. and Yang, R. and Zhang, W. and Liu, J. Y. and Xiao, H. and Chen, G. F. and Taylor, A. J. and Yarotski, D. A. and Prasankumar, R. P. and Qiu, X. G.},
  journal = {Phys. Rev. B},
  volume = {93},
  issue = {12},
  pages = {121110(R)},
  numpages = {5},
  year = {2016},
  month = {Mar},
  publisher = {American Physical Society},
  doi = {10.1103/PhysRevB.93.121110},
  url = {https://link.aps.org/doi/10.1103/PhysRevB.93.121110}
}

@article{Op_conduc_Dirac_Weyl4,
author = {Pronin, Artem V. and Dressel, Martin},
title = {Nodal Semimetals: A Survey on Optical Conductivity},
journal = {physica status solidi (b)},
volume = {258},
number = {1},
pages = {2000027},
doi = {https://doi.org/10.1002/pssb.202000027},
url = {https://onlinelibrary.wiley.com/doi/abs/10.1002/pssb.202000027},
year = {2021}
}

@article{Op_conduc_Dirac_Weyl5,
  title = {Optical and transport properties in three-dimensional Dirac and Weyl semimetals},
  author = {Tabert, C. J. and Carbotte, J. P. and Nicol, E. J.},
  journal = {Phys. Rev. B},
  volume = {93},
  issue = {8},
  pages = {085426},
  numpages = {18},
  year = {2016},
  month = {Feb},
  publisher = {American Physical Society},
  doi = {10.1103/PhysRevB.93.085426},
  url = {https://link.aps.org/doi/10.1103/PhysRevB.93.085426}
}

@article{Op_conduc_Dirac_Weyl6,
  title = {Optical conductivity of multi-Weyl semimetals},
  author = {Ahn, Seongjin and Mele, E. J. and Min, Hongki},
  journal = {Phys. Rev. B},
  volume = {95},
  issue = {16},
  pages = {161112(R)},
  numpages = {5},
  year = {2017},
  month = {Apr},
  publisher = {American Physical Society},
  doi = {10.1103/PhysRevB.95.161112},
  url = {https://link.aps.org/doi/10.1103/PhysRevB.95.161112}
}

@Article{Op_graphene1,
author={Li, Z. Q.
and Henriksen, E. A.
and Jiang, Z.
and Hao, Z.
and Martin, M. C.
and Kim, P.
and Stormer, H. L.
and Basov, D. N.},
title={Dirac charge dynamics in graphene by infrared spectroscopy},
journal={Nature Physics},
year={2008},
month={Jul},
day={01},
volume={4},
number={7},
pages={532-535},
issn={1745-2481},
doi={10.1038/nphys989},
url={https://doi.org/10.1038/nphys989}
}

@article{Op_graphene2,
  title = {Effect of electron-phonon interaction on spectroscopies in graphene},
  author = {Carbotte, J. P. and Nicol, E. J. and Sharapov, S. G.},
  journal = {Phys. Rev. B},
  volume = {81},
  issue = {4},
  pages = {045419},
  numpages = {19},
  year = {2010},
  month = {Jan},
  publisher = {American Physical Society},
  doi = {10.1103/PhysRevB.81.045419},
  url = {https://link.aps.org/doi/10.1103/PhysRevB.81.045419}
}

@article{Op_TI,
  title = {Landau level spectroscopy of surface states in the topological insulator Bi${}_{0.91}$Sb${}_{0.09}$ via magneto-optics},
  author = {Schafgans, A. A. and Post, K. W. and Taskin, A. A. and Ando, Yoichi and Qi, Xiao-Liang and Chapler, B. C. and Basov, D. N.},
  journal = {Phys. Rev. B},
  volume = {85},
  issue = {19},
  pages = {195440},
  numpages = {6},
  year = {2012},
  month = {May},
  publisher = {American Physical Society},
  doi = {10.1103/PhysRevB.85.195440},
  url = {https://link.aps.org/doi/10.1103/PhysRevB.85.195440}
}

@Article{NLHC_Ex1,
author={Du, Z. Z.
and Lu, Hai-Zhou
and Xie, X. C.},
title={Nonlinear Hall effects},
journal={Nature Reviews Physics},
year={2021},
month={Nov},
day={01},
volume={3},
number={11},
pages={744-752},
issn={2522-5820},
doi={10.1038/s42254-021-00359-6},
url={https://doi.org/10.1038/s42254-021-00359-6}
}

@article{NLHC_Ex2,
title = {Non-linear Hall effects: Mechanisms and materials},
journal = {Materials Today Electronics},
volume = {8},
pages = {100101},
year = {2024},
issn = {2772-9494},
doi = {https://doi.org/10.1016/j.mtelec.2024.100101},
url = {https://www.sciencedirect.com/science/article/pii/S2772949424000135},
author = {Arka Bandyopadhyay and Nesta Benno Joseph and Awadhesh Narayan}
}

@Article{NLHC_Ex3,
AUTHOR = {Shuo Wang and Wei Niu and Yue-Wen Fang},
TITLE = {Nonlinear Hall effect in two-dimensional materials},
JOURNAL = {Microstructures},
VOLUME = {5},
YEAR = {2025},
NUMBER = {3},
ARTICLE-NUMBER = {2025060},
URL = {https://www.oaepublish.com/articles/microstructures.2024.129},
ISSN = {2770-2995},
DOI = {10.20517/microstructures.2024.129}
}

@article{NLHC_Ex5,
  title = {Band Signatures for Strong Nonlinear Hall Effect in Bilayer ${\mathrm{WTe}}_{2}$},
  author = {Du, Z. Z. and Wang, C. M. and Lu, Hai-Zhou and Xie, X. C.},
  journal = {Phys. Rev. Lett.},
  volume = {121},
  issue = {26},
  pages = {266601},
  numpages = {6},
  year = {2018},
  month = {Dec},
  publisher = {American Physical Society},
  doi = {10.1103/PhysRevLett.121.266601},
  url = {https://link.aps.org/doi/10.1103/PhysRevLett.121.266601}
}

@Article{NLHC_Ex6,
author={Du, Z. Z.
and Wang, C. M.
and Li, Shuai
and Lu, Hai-Zhou
and Xie, X. C.},
title={Disorder-induced nonlinear Hall effect with time-reversal symmetry},
journal={Nature Communications},
year={2019},
month={Jul},
day={11},
volume={10},
number={1},
pages={3047},
issn={2041-1723},
doi={10.1038/s41467-019-10941-3},
url={https://doi.org/10.1038/s41467-019-10941-3}
}

@article{NLHC_In1,
  title = {Field Induced Positional Shift of Bloch Electrons and Its Dynamical Implications},
  author = {Gao, Yang and Yang, Shengyuan A. and Niu, Qian},
  journal = {Phys. Rev. Lett.},
  volume = {112},
  issue = {16},
  pages = {166601},
  numpages = {5},
  year = {2014},
  month = {Apr},
  publisher = {American Physical Society},
  doi = {10.1103/PhysRevLett.112.166601},
  url = {https://link.aps.org/doi/10.1103/PhysRevLett.112.166601}
}

@article{NLHC_In2,
  title = {Band structure engineering of ideal fractional Chern insulators},
  author = {Lee, Ching Hua and Claassen, Martin and Thomale, Ronny},
  journal = {Phys. Rev. B},
  volume = {96},
  issue = {16},
  pages = {165150},
  numpages = {16},
  year = {2017},
  month = {Oct},
  publisher = {American Physical Society},
  doi = {10.1103/PhysRevB.96.165150},
  url = {https://link.aps.org/doi/10.1103/PhysRevB.96.165150}
}

@article{NLHC_In3,
  title = {Intrinsic Nonlinear Planar Hall Effect},
  author = {Huang, Yue-Xin and Feng, Xiaolong and Wang, Hui and Xiao, Cong and Yang, Shengyuan A.},
  journal = {Phys. Rev. Lett.},
  volume = {130},
  issue = {12},
  pages = {126303},
  numpages = {6},
  year = {2023},
  month = {Mar},
  publisher = {American Physical Society},
  doi = {10.1103/PhysRevLett.130.126303},
  url = {https://link.aps.org/doi/10.1103/PhysRevLett.130.126303}
}

@article{NLHC_In4,
doi={10.1126/science.adf1506},
author = {Anyuan Gao  and Yu-Fei Liu  and Jian-Xiang Qiu  and Barun Ghosh  and Thaís V. Trevisan  and Yugo Onishi  and Chaowei Hu  and Tiema Qian  and Hung-Ju Tien  and Shao-Wen Chen  and Mengqi Huang  and Damien Bérubé  and Houchen Li  and Christian Tzschaschel  and Thao Dinh  and Zhe Sun  and Sheng-Chin Ho  and Shang-Wei Lien  and Bahadur Singh  and Kenji Watanabe  and Takashi Taniguchi  and David C. Bell  and Hsin Lin  and Tay-Rong Chang  and Chunhui Rita Du  and Arun Bansil  and Liang Fu  and Ni Ni  and Peter P. Orth  and Qiong Ma  and Su-Yang Xu },
title = {Quantum metric nonlinear Hall effect in a topological antiferromagnetic heterostructure},
journal = {Science},
volume = {381},
number = {6654},
pages = {181-186},
year = {2023},
doi = {10.1126/science.adf1506},
URL = {https://www.science.org/doi/abs/10.1126/science.adf1506}
}

@article{NLHC_In5,
  title = {Unification of Nonlinear Anomalous Hall Effect and Nonreciprocal Magnetoresistance in Metals by the Quantum Geometry},
  author = {Kaplan, Daniel and Holder, Tobias and Yan, Binghai},
  journal = {Phys. Rev. Lett.},
  volume = {132},
  issue = {2},
  pages = {026301},
  numpages = {7},
  year = {2024},
  month = {Jan},
  publisher = {American Physical Society},
  doi = {10.1103/PhysRevLett.132.026301},
  url = {https://link.aps.org/doi/10.1103/PhysRevLett.132.026301}
}

@article{NLHC_In6,
  title = {Intrinsic Second-Order Anomalous Hall Effect and Its Application in Compensated Antiferromagnets},
  author = {Liu, Huiying and Zhao, Jianzhou and Huang, Yue-Xin and Wu, Weikang and Sheng, Xian-Lei and Xiao, Cong and Yang, Shengyuan A.},
  journal = {Phys. Rev. Lett.},
  volume = {127},
  issue = {27},
  pages = {277202},
  numpages = {6},
  year = {2021},
  month = {Dec},
  publisher = {American Physical Society},
  doi = {10.1103/PhysRevLett.127.277202},
  url = {https://link.aps.org/doi/10.1103/PhysRevLett.127.277202}
}

@article{NLHC_In7,
  title = {Magnetic Field Induced Quantum Metric Dipole in Dirac Semimetal ${\mathrm{Cd}}_{3}{\mathrm{As}}_{2}$},
  author = {Zhao, Tong-Yang and Wang, An-Qi and Zhang, Zhen-Tao and Cao, Zheng-Yang and Liu, Xing-Yu and Liao, Zhi-Min},
  journal = {Phys. Rev. Lett.},
  volume = {135},
  issue = {2},
  pages = {026601},
  numpages = {8},
  year = {2025},
  month = {Jul},
  publisher = {American Physical Society},
  doi = {10.1103/bzdt-yxk2},
  url = {https://link.aps.org/doi/10.1103/bzdt-yxk2}
}

@article{NLHC_QMD,
  title = {Intrinsic nonlinear conductivity induced by quantum geometry in altermagnets and measurement of the in-plane N\'eel vector},
  author = {Ezawa, Motohiko},
  journal = {Phys. Rev. B},
  volume = {110},
  issue = {24},
  pages = {L241405},
  numpages = {6},
  year = {2024},
  month = {Dec},
  publisher = {American Physical Society},
  doi = {10.1103/PhysRevB.110.L241405},
  url = {https://link.aps.org/doi/10.1103/PhysRevB.110.L241405}
}

@article{Weyl_SM1,
  title = {Topological semimetal and Fermi-arc surface states in the electronic structure of pyrochlore iridates},
  author = {Wan, Xiangang and Turner, Ari M. and Vishwanath, Ashvin and Savrasov, Sergey Y.},
  journal = {Phys. Rev. B},
  volume = {83},
  issue = {20},
  pages = {205101},
  numpages = {9},
  year = {2011},
  month = {May},
  publisher = {American Physical Society},
  doi = {10.1103/PhysRevB.83.205101},
  url = {https://link.aps.org/doi/10.1103/PhysRevB.83.205101}
}

@article{Weyl_SM2,
  title = {Weyl Semimetal Phase in Noncentrosymmetric Transition-Metal Monophosphides},
  author = {Weng, Hongming and Fang, Chen and Fang, Zhong and Bernevig, B. Andrei and Dai, Xi},
  journal = {Phys. Rev. X},
  volume = {5},
  issue = {1},
  pages = {011029},
  numpages = {10},
  year = {2015},
  month = {Mar},
  publisher = {American Physical Society},
  doi = {10.1103/PhysRevX.5.011029},
  url = {https://link.aps.org/doi/10.1103/PhysRevX.5.011029}
}

@article{Weyl_SM3,
  title = {Experimental Discovery of Weyl Semimetal TaAs},
  author = {Lv, B. Q. and Weng, H. M. and Fu, B. B. and Wang, X. P. and Miao, H. and Ma, J. and Richard, P. and Huang, X. C. and Zhao, L. X. and Chen, G. F. and Fang, Z. and Dai, X. and Qian, T. and Ding, H.},
  journal = {Phys. Rev. X},
  volume = {5},
  issue = {3},
  pages = {031013},
  numpages = {8},
  year = {2015},
  month = {Jul},
  publisher = {American Physical Society},
  doi = {10.1103/PhysRevX.5.031013},
  url = {https://link.aps.org/doi/10.1103/PhysRevX.5.031013}
}

@Article{Weyl_SM4,
author={Huang, Shin-Ming
and Xu, Su-Yang
and Belopolski, Ilya
and Lee, Chi-Cheng
and Chang, Guoqing
and Wang, BaoKai
and Alidoust, Nasser
and Bian, Guang
and Neupane, Madhab
and Zhang, Chenglong
and Jia, Shuang
and Bansil, Arun
and Lin, Hsin
and Hasan, M. Zahid},
title={A Weyl Fermion semimetal with surface Fermi arcs in the transition metal monopnictide TaAs class},
journal={Nature Communications},
year={2015},
month={Jun},
day={12},
volume={6},
number={1},
pages={7373},
issn={2041-1723},
doi={10.1038/ncomms8373},
url={https://doi.org/10.1038/ncomms8373}
}

@Article{Weyl_SM5,
author={Shekhar, Chandra
and Nayak, Ajaya K.
and Sun, Yan
and Schmidt, Marcus
and Nicklas, Michael
and Leermakers, Inge
and Zeitler, Uli
and Skourski, Yurii
and Wosnitza, Jochen
and Liu, Zhongkai
and Chen, Yulin
and Schnelle, Walter
and Borrmann, Horst
and Grin, Yuri
and Felser, Claudia
and Yan, Binghai},
title={Extremely large magnetoresistance and ultrahigh mobility in the topological Weyl semimetal candidate NbP},
journal={Nature Physics},
year={2015},
month={Aug},
day={01},
volume={11},
number={8},
pages={645-649},
issn={1745-2481},
doi={10.1038/nphys3372},
url={https://doi.org/10.1038/nphys3372}
}

@Article{Weyl_SM6,
author={Xu, Su-Yang
and Alidoust, Nasser
and Belopolski, Ilya
and Yuan, Zhujun
and Bian, Guang
and Chang, Tay-Rong
and Zheng, Hao
and Strocov, Vladimir N.
and Sanchez, Daniel S.
and Chang, Guoqing
and Zhang, Chenglong
and Mou, Daixiang
and Wu, Yun
and Huang, Lunan
and Lee, Chi-Cheng
and Huang, Shin-Ming
and Wang, BaoKai
and Bansil, Arun
and Jeng, Horng-Tay
and Neupert, Titus
and Kaminski, Adam
and Lin, Hsin
and Jia, Shuang
and Zahid Hasan, M.},
title={Discovery of a Weyl fermion state with Fermi arcs in niobium arsenide},
journal={Nature Physics},
year={2015},
month={Sep},
day={01},
volume={11},
number={9},
pages={748-754},
issn={1745-2481},
doi={10.1038/nphys3437},
url={https://doi.org/10.1038/nphys3437}
}

@article{Weyl_SM7,
doi={10.1126/science.aaa9273},
author = {Ling Lu  and Zhiyu Wang  and Dexin Ye  and Lixin Ran  and Liang Fu  and John D. Joannopoulos  and Marin Soljačić },
title = {Experimental observation of Weyl points},
journal = {Science},
volume = {349},
number = {6248},
pages = {622-624},
year = {2015},
doi = {10.1126/science.aaa9273},
URL = {https://www.science.org/doi/abs/10.1126/science.aaa9273}
}

@article{Weyl_SM8,
doi={10.1126/science.1256742},
author = {Su-Yang Xu  and Chang Liu  and Satya K. Kushwaha  and Raman Sankar  and Jason W. Krizan  and Ilya Belopolski  and Madhab Neupane  and Guang Bian  and Nasser Alidoust  and Tay-Rong Chang  and Horng-Tay Jeng  and Cheng-Yi Huang  and Wei-Feng Tsai  and Hsin Lin  and Pavel P. Shibayev  and Fang-Cheng Chou  and Robert J. Cava  and M. Zahid Hasan },
title = {Observation of Fermi arc surface states in a topological metal},
journal = {Science},
volume = {347},
number = {6219},
pages = {294-298},
year = {2015},
doi = {10.1126/science.1256742},
URL = {https://www.science.org/doi/abs/10.1126/science.1256742},
}

@article{Dirac_SM1,
  title = {Dirac semimetal and topological phase transitions in ${A}_{3}$Bi ($A=\text{Na}$, K, Rb)},
  author = {Wang, Zhijun and Sun, Yan and Chen, Xing-Qiu and Franchini, Cesare and Xu, Gang and Weng, Hongming and Dai, Xi and Fang, Zhong},
  journal = {Phys. Rev. B},
  volume = {85},
  issue = {19},
  pages = {195320},
  numpages = {5},
  year = {2012},
  month = {May},
  publisher = {American Physical Society},
  doi = {10.1103/PhysRevB.85.195320},
  url = {https://link.aps.org/doi/10.1103/PhysRevB.85.195320}
}

@article{Dirac_SM2,
  title = {Three-dimensional Dirac semimetal and quantum transport in Cd${}_{3}$As${}_{2}$},
  author = {Wang, Zhijun and Weng, Hongming and Wu, Quansheng and Dai, Xi and Fang, Zhong},
  journal = {Phys. Rev. B},
  volume = {88},
  issue = {12},
  pages = {125427},
  numpages = {6},
  year = {2013},
  month = {Sep},
  publisher = {American Physical Society},
  doi = {10.1103/PhysRevB.88.125427},
  url = {https://link.aps.org/doi/10.1103/PhysRevB.88.125427}
}

@article{Dirac_SM3,
  title = {Experimental Realization of a Three-Dimensional Dirac Semimetal},
  author = {Borisenko, Sergey and Gibson, Quinn and Evtushinsky, Danil and Zabolotnyy, Volodymyr and B\"uchner, Bernd and Cava, Robert J.},
  journal = {Phys. Rev. Lett.},
  volume = {113},
  issue = {2},
  pages = {027603},
  numpages = {5},
  year = {2014},
  month = {Jul},
  publisher = {American Physical Society},
  doi = {10.1103/PhysRevLett.113.027603},
  url = {https://link.aps.org/doi/10.1103/PhysRevLett.113.027603}
}

@article{Dirac_SM4,
doi={10.1126/science.1245085},
author = {Z. K. Liu  and B. Zhou  and Y. Zhang  and Z. J. Wang  and H. M. Weng  and D. Prabhakaran  and S.-K. Mo  and Z. X. Shen  and Z. Fang  and X. Dai  and Z. Hussain  and Y. L. Chen },
title = {Discovery of a Three-Dimensional Topological Dirac Semimetal, $Na_{3}Bi$},
journal = {Science},
volume = {343},
number = {6173},
pages = {864-867},
year = {2014},
doi = {10.1126/science.1245085},
URL = {https://www.science.org/doi/abs/10.1126/science.1245085}
}

@Article{Dirac_SM5,
author={Liu, Z. K.
and Jiang, J.
and Zhou, B.
and Wang, Z. J.
and Zhang, Y.
and Weng, H. M.
and Prabhakaran, D.
and Mo, S.-K.
and Peng, H.
and Dudin, P.
and Kim, T.
and Hoesch, M.
and Fang, Z.
and Dai, X.
and Shen, Z. X.
and Feng, D. L.
and Hussain, Z.
and Chen, Y. L.},
title={A stable three-dimensional topological Dirac semimetal Cd3As2},
journal={Nature Materials},
year={2014},
month={Jul},
day={01},
volume={13},
number={7},
pages={677-681},
issn={1476-4660},
doi={10.1038/nmat3990},
url={https://doi.org/10.1038/nmat3990}
}

@article{NLSM1,
  title = {Topological nodal semimetals},
  author = {Burkov, A. A. and Hook, M. D. and Balents, Leon},
  journal = {Phys. Rev. B},
  volume = {84},
  issue = {23},
  pages = {235126},
  numpages = {14},
  year = {2011},
  month = {Dec},
  publisher = {American Physical Society},
  doi = {10.1103/PhysRevB.84.235126},
  url = {https://link.aps.org/doi/10.1103/PhysRevB.84.235126}
}

@article{NLSM2,
  title = {Tunable line node semimetals},
  author = {Phillips, Michael and Aji, Vivek},
  journal = {Phys. Rev. B},
  volume = {90},
  issue = {11},
  pages = {115111},
  numpages = {7},
  year = {2014},
  month = {Sep},
  publisher = {American Physical Society},
  doi = {10.1103/PhysRevB.90.115111},
  url = {https://link.aps.org/doi/10.1103/PhysRevB.90.115111}
}

@article{NLSM3,
  title = {Topological node-line semimetal in three-dimensional graphene networks},
  author = {Weng, Hongming and Liang, Yunye and Xu, Qiunan and Yu, Rui and Fang, Zhong and Dai, Xi and Kawazoe, Yoshiyuki},
  journal = {Phys. Rev. B},
  volume = {92},
  issue = {4},
  pages = {045108},
  numpages = {8},
  year = {2015},
  month = {Jul},
  publisher = {American Physical Society},
  doi = {10.1103/PhysRevB.92.045108},
  url = {https://link.aps.org/doi/10.1103/PhysRevB.92.045108}
}

@article{NLSM4,
  title = {Topological nodal line semimetals with and without spin-orbital coupling},
  author = {Fang, Chen and Chen, Yige and Kee, Hae-Young and Fu, Liang},
  journal = {Phys. Rev. B},
  volume = {92},
  issue = {8},
  pages = {081201(R)},
  numpages = {5},
  year = {2015},
  month = {Aug},
  publisher = {American Physical Society},
  doi = {10.1103/PhysRevB.92.081201},
  url = {https://link.aps.org/doi/10.1103/PhysRevB.92.081201}
}

@Article{NLSM5,
author={Chen, Yige
and Lu, Yuan-Ming
and Kee, Hae-Young},
title={Topological crystalline metal in orthorhombic perovskite iridates},
journal={Nature Communications},
year={2015},
month={Mar},
day={16},
volume={6},
number={1},
pages={6593},
issn={2041-1723},
doi={10.1038/ncomms7593},
url={https://doi.org/10.1038/ncomms7593}
}

@article{NLSM6,
doi={10.7566/JPSJ.85.013708},
author = {Yamakage ,Ai and Yamakawa ,Youichi and Tanaka ,Yukio and Okamoto ,Yoshihiko},
title = {Line-Node Dirac Semimetal and Topological Insulating Phase in Noncentrosymmetric Pnictides CaAgX (X = P, As)},
journal = {Journal of the Physical Society of Japan},
volume = {85},
number = {1},
pages = {013708},
year = {2016},
doi = {10.7566/JPSJ.85.013708},
URL = {https://doi.org/10.7566/JPSJ.85.013708},
}

@article{NLSM7,
  title = {${\mathrm{Ca}}_{3}{\mathrm{P}}_{2}$ and other topological semimetals with line nodes and drumhead surface states},
  author = {Chan, Y.-H. and Chiu, Ching-Kai and Chou, M. Y. and Schnyder, Andreas P.},
  journal = {Phys. Rev. B},
  volume = {93},
  issue = {20},
  pages = {205132},
  numpages = {16},
  year = {2016},
  month = {May},
  publisher = {American Physical Society},
  doi = {10.1103/PhysRevB.93.205132},
  url = {https://link.aps.org/doi/10.1103/PhysRevB.93.205132}
}

@article{DTNL1,
  title = {Dirac Line Nodes in Inversion-Symmetric Crystals},
  author = {Kim, Youngkuk and Wieder, Benjamin J. and Kane, C. L. and Rappe, Andrew M.},
  journal = {Phys. Rev. Lett.},
  volume = {115},
  issue = {3},
  pages = {036806},
  numpages = {5},
  year = {2015},
  month = {Jul},
  publisher = {American Physical Society},
  doi = {10.1103/PhysRevLett.115.036806},
  url = {https://link.aps.org/doi/10.1103/PhysRevLett.115.036806}
}

@article{DTNL2,
  title = {Topological Node-Line Semimetal and Dirac Semimetal State in Antiperovskite ${\mathrm{Cu}}_{3}\mathrm{PdN}$},
  author = {Yu, Rui and Weng, Hongming and Fang, Zhong and Dai, Xi and Hu, Xiao},
  journal = {Phys. Rev. Lett.},
  volume = {115},
  issue = {3},
  pages = {036807},
  numpages = {5},
  year = {2015},
  month = {Jul},
  publisher = {American Physical Society},
  doi = {10.1103/PhysRevLett.115.036807},
  url = {https://link.aps.org/doi/10.1103/PhysRevLett.115.036807}
}

@Article{DTNL3,
author={Bzdu{\v{s}}ek, Tom{\'a}{\v{s}}
and Wu, QuanSheng
and R{\"u}egg, Andreas
and Sigrist, Manfred
and Soluyanov, Alexey A.},
title={Nodal-chain metals},
journal={Nature},
year={2016},
month={Oct},
day={01},
volume={538},
number={7623},
pages={75-78},
issn={1476-4687},
doi={10.1038/nature19099},
url={https://doi.org/10.1038/nature19099}
}

@article{DTNL4,
  title = {From Nodal Chain Semimetal to Weyl Semimetal in HfC},
  author = {Yu, Rui and Wu, Quansheng and Fang, Zhong and Weng, Hongming},
  journal = {Phys. Rev. Lett.},
  volume = {119},
  issue = {3},
  pages = {036401},
  numpages = {5},
  year = {2017},
  month = {Jul},
  publisher = {American Physical Society},
  doi = {10.1103/PhysRevLett.119.036401},
  url = {https://link.aps.org/doi/10.1103/PhysRevLett.119.036401}
}

@article{Hopf1,
  title = {Nodal-link semimetals},
  author = {Yan, Zhongbo and Bi, Ren and Shen, Huitao and Lu, Ling and Zhang, Shou-Cheng and Wang, Zhong},
  journal = {Phys. Rev. B},
  volume = {96},
  issue = {4},
  pages = {041103(R)},
  numpages = {8},
  year = {2017},
  month = {Jul},
  publisher = {American Physical Society},
  doi = {10.1103/PhysRevB.96.041103},
  url = {https://link.aps.org/doi/10.1103/PhysRevB.96.041103}
}

@article{Hopf2,
  title = {Topological semimetals with a double-helix nodal link},
  author = {Chen, Wei and Lu, Hai-Zhou and Hou, Jing-Min},
  journal = {Phys. Rev. B},
  volume = {96},
  issue = {4},
  pages = {041102(R)},
  numpages = {5},
  year = {2017},
  month = {Jul},
  publisher = {American Physical Society},
  doi = {10.1103/PhysRevB.96.041102},
  url = {https://link.aps.org/doi/10.1103/PhysRevB.96.041102}
}

@article{Hopf3,
  title = {Topological semimetals carrying arbitrary Hopf numbers: Fermi surface topologies of a Hopf link, Solomon's knot, trefoil knot, and other linked nodal varieties},
  author = {Ezawa, Motohiko},
  journal = {Phys. Rev. B},
  volume = {96},
  issue = {4},
  pages = {041202(R)},
  numpages = {5},
  year = {2017},
  month = {Jul},
  publisher = {American Physical Society},
  doi = {10.1103/PhysRevB.96.041202},
  url = {https://link.aps.org/doi/10.1103/PhysRevB.96.041202}
}

@article{Hopf4,
  title = {Quantum oscillation in Hopf-link semimetals},
  author = {Shi, Lei and Liu, Xiaoxiong and Wang, C. M. and Liu, Tianyu and Lu, Hai-Zhou and Xie, X. C.},
  journal = {Phys. Rev. B},
  volume = {111},
  issue = {20},
  pages = {L201103},
  numpages = {6},
  year = {2025},
  month = {May},
  publisher = {American Physical Society},
  doi = {10.1103/PhysRevB.111.L201103},
  url = {https://link.aps.org/doi/10.1103/PhysRevB.111.L201103}
}

@article{Knot1,
  title = {Nodal-knot semimetals},
  author = {Bi, Ren and Yan, Zhongbo and Lu, Ling and Wang, Zhong},
  journal = {Phys. Rev. B},
  volume = {96},
  issue = {20},
  pages = {201305(R)},
  numpages = {7},
  year = {2017},
  month = {Nov},
  publisher = {American Physical Society},
  doi = {10.1103/PhysRevB.96.201305},
  url = {https://link.aps.org/doi/10.1103/PhysRevB.96.201305}
}

@article{Knot2,
  title = {Jones Polynomial and Knot Transitions in Hermitian and non-Hermitian Topological Semimetals},
  author = {Yang, Zhesen and Chiu, Ching-Kai and Fang, Chen and Hu, Jiangping},
  journal = {Phys. Rev. Lett.},
  volume = {124},
  issue = {18},
  pages = {186402},
  numpages = {6},
  year = {2020},
  month = {May},
  publisher = {American Physical Society},
  doi = {10.1103/PhysRevLett.124.186402},
  url = {https://link.aps.org/doi/10.1103/PhysRevLett.124.186402}
}

@Article{NHLC_expt,
author={Wang, Naizhou
and Kaplan, Daniel
and Zhang, Zhaowei
and Holder, Tobias
and Cao, Ning
and Wang, Aifeng
and Zhou, Xiaoyuan
and Zhou, Feifei
and Jiang, Zhengzhi
and Zhang, Chusheng
and Ru, Shihao
and Cai, Hongbing
and Watanabe, Kenji
and Taniguchi, Takashi
and Yan, Binghai
and Gao, Weibo},
title={Quantum-metric-induced nonlinear transport in a topological antiferromagnet},
journal={Nature},
year={2023},
month={Sep},
day={01},
volume={621},
number={7979},
pages={487-492},
issn={1476-4687},
doi={10.1038/s41586-023-06363-3},
url={https://doi.org/10.1038/s41586-023-06363-3}
}

@misc{supp,
  title = {},
  author = {},
  year = {},
  note = {See the Supplemental Material (SM) for a detailed discussion and derivation of the low-energy continuum model from the lattice model of the Hopf-link semimetal, as well as for the expressions of various quantum metric components of the Hopf-link semimetal.},
  url = {}
}

\normalsize\clearpage
\begin{onecolumngrid}
	\begin{center}
		{\fontsize{12}{12}\selectfont
			\textbf{Supplemental Material for ``Quantum Geometric Signatures of Link-Unlink Transitions and Nonlinear Hall Response in Hopf-Link Semimetals''\\[5mm]}}
		{\normalsize  Kamalesh Bera\orcidA{},$^{1,2,*}$ Arijit Saha\orcidC{},$^{1,2,*}$ Debashree Chowdhury \orcidB{},$^{3}$ \\[1mm]}
		{\small $^1$\textit{Institute of Physics, Sachivalaya Marg, Bhubaneswar-751005, India}\\[0.5mm]}
		{\small $^2$\textit{Homi Bhabha National Institute, Training School Complex, Anushakti Nagar, Mumbai 400094, India}\\[0.5mm]}
		{\small $^3$\textit{Centre for Nanotechnology, IIT Roorkee, Roorkee, Uttarakhand 247667, India}\\[0.5mm]}
	\end{center}
	
	\newcounter{defcounter}
	\setcounter{defcounter}{0}
	\setcounter{equation}{0}
	\renewcommand{\theequation}{S\arabic{equation}}
	\setcounter{figure}{0}
	\renewcommand{\thefigure}{S\arabic{figure}}
	\setcounter{page}{1}
	\pagenumbering{roman}
	\renewcommand{\thesection}{S\arabic{section}}
	
\vspace{0.3cm}	

This supplementary text provides couple of analysis that complements the main manuscript. In Sec.~\ref{AppA}, 
we describe the lattice to continuum model conversion of the Hopf-Link Semimetal. Sec.~\ref{AppB} is devoted
to the discussion of quantum metric components of Hopf-Link Semimetal.

\section{From Lattice to Continuum model of the Hopf-Link Semimetal} \label{AppA}
In an earlier work on the construction of Hopf semimetals~\cite{Hopf1}, two lattice models were proposed based on the relevant symmetry constraints. Here, we revisit these two models and present their corresponding Hopf-linked phases.

The first model Hamiltonian for the Hopf semimetal is given by
\begin{eqnarray}
	H_{XZ}(\mathbf{k}) &=&
	\Bigg[2\sin{k_x}\sin{k_z}+2\sin{k_y}\left(
	\sum_{i=x,y,z}\cos{k_i}-m_0\right)\Bigg]\sigma_x
	\nonumber\\
	&+&
	\Bigg[\sin^2{k_x}+\sin^2{k_y}-\sin^2{k_z}+\left(
	\sum_{i=x,y,z}\cos{k_i}-m_0\right)\Bigg]\sigma_z \ ,
	\label{TB_Hamiltonian1}
\end{eqnarray}
Here, $k_x$, $k_y$, and $k_z$ denote the crystal momenta, while $m_0$ represents the Hopf mass parameter in the lattice model. The matrices $\sigma_x$ and $\sigma_z$ are the Pauli matrices in the pseudospin space. The nodal link is formed when the coefficients of both $\sigma_x$ and $\sigma_z$ simultaneously vanish. 
In Fig.~\ref{Hopf-link_lattice}(a), we depict the resulting linked phase for $m_0=2.5$, where the nodal structure consists of two linked rings.

The second model Hamiltonian for the Hopf semimetal is given by
\begin{eqnarray}
	H_{XY}(\mathbf{k}) &=&
	\Bigg[\sin{k_x}\sin{k_z}+\sin{k_y}
	\left(\sum_{i=x,y,z}\cos{k_i}-m_0\right)
	\Bigg]\sigma_x
	\nonumber\\
	&+&
	\Bigg[-\sin{k_y}\sin{k_z}+\sin{k_x}
	\left(\sum_{i=x,y,z}\cos{k_i}-m_0\right)
	\Bigg]\sigma_y \ ,
	\label{TB_Hamiltonian2}
\end{eqnarray}
This Hamiltonian also exhibits a nodal link, as shown in Fig.~\ref{Hopf-link_lattice}(b), considering the same value of the Hopf mass parameter $m_0$. For computational convenience, we chose this model Hamiltonian and derive the corresponding low-energy effective continuum model below.

\begin{figure}[h]
	\subfigure{\includegraphics[width=0.41\textwidth]{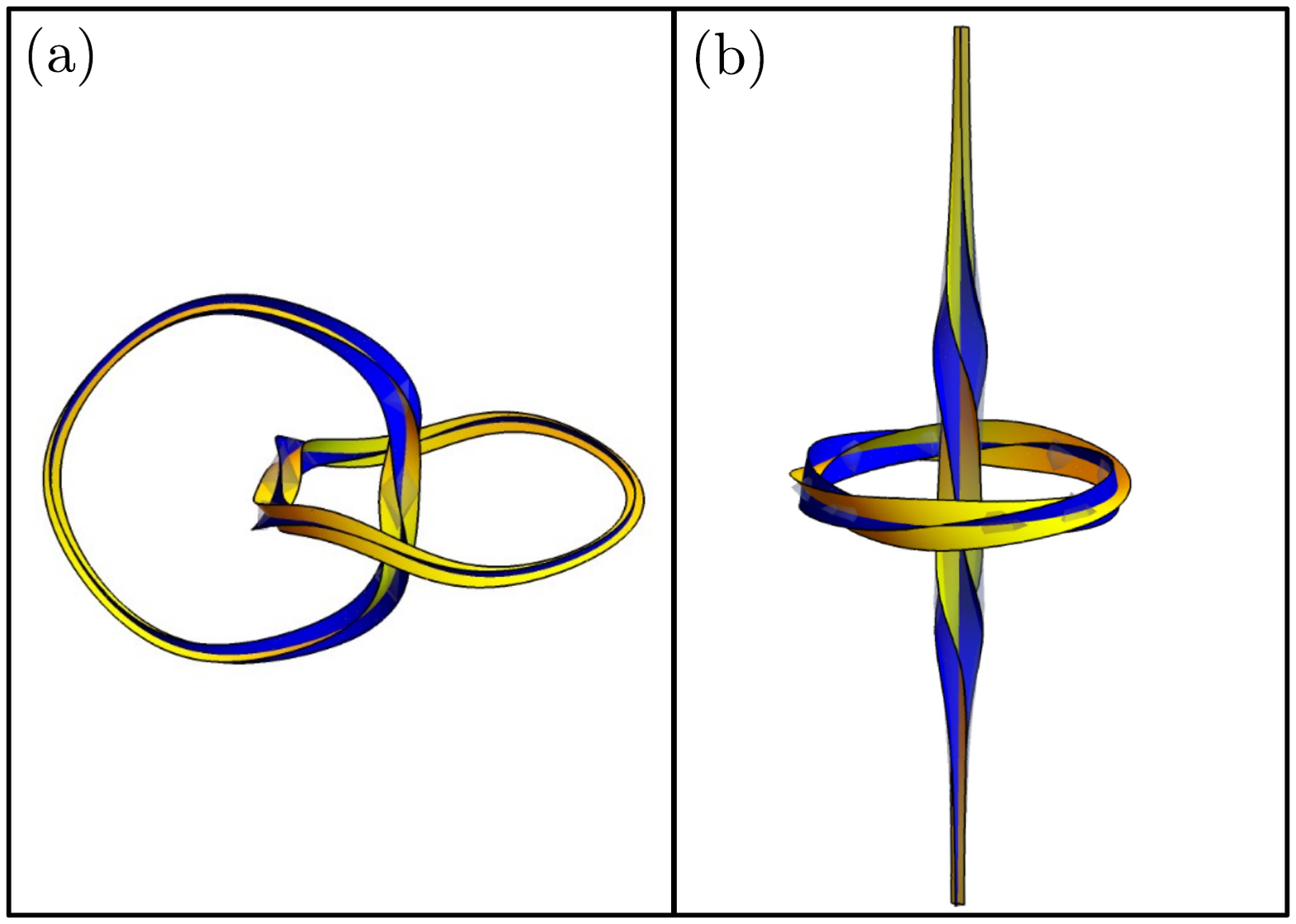}}
	\caption{Linked phases corresponding to the Hopf semimetal, obtained from two different lattice model Hamiltonians $H_{XZ}$ and $H_{XY}$, are displayed in panels (a) and (b), respectively. In both the cases, we choose the Hopf mass parameter as $m_0 = 2.5$.}
	\label{Hopf-link_lattice}
\end{figure}

At low energies, we use the approximations $\sin{k_i}\approx k_i$ and $\cos{k_i}\approx 1-k_i^2/2$ to construct the corresponding continuum model. Thus, near the critical point, the lattice Hamiltonian $H_{XY}$ [Eq.~(\ref{TB_Hamiltonian2})] can be expanded to leading order in momentum ($k_x$, $k_y$) as
\begin{eqnarray}
	H(\mathbf{k}) &=
	\Big[
	(m-Ck^2)k_y+k_xk_z
	\Big]\sigma_x
	+
	\Big[
	(m-Ck^2)k_x-k_yk_z
	\Big]\sigma_y \ ,
	\label{Continuum_H}
\end{eqnarray}
where, $m=3-m_0$, $C=0.5$, and $k^2=k_x^2+k_y^2+k_z^2.$
As discussed in the main text, all the analyses presented in this work are based on this low-energy continuum model.

\section{Quantum metric components of Hopf-link semimetal} \label{AppB}
In the main text, we focus on the $zz$-component of the quantum metric in the $k_x$--$k_y$ plane. For completeness, here we present additional components of the quantum metric in both the $k_x$--$k_y$ and $k_x$--$k_z$ planes to provide a more comprehensive understanding of the quantum geometric properties of the Hopf-link semimetal. Below we write the analytical expressions for the quantum metric components and present their behavior in Fig.~\ref{QMs} subsequently,

\begin{eqnarray}
g_{xx}
&=&
\frac{
\left[ -2C k_x k_{s}^{2} k_z + C^2 k_y k_r^4
	-2 C k_y k_r^2m + k_y M_z \right]^2
}{
	k_{s}^{4}
	\left[
	k_z^2+\left(- C k_r^2+m\right)^2
	\right]^2
}\ ,\\
g_{xy}
&=&-\frac{
	\begin{aligned}
		\left[
		2 C k_y k_{s}^{2} k_z + C^2 k_x k_r^4
		-2 C k_x k_r^2 m +k_x M_z
		\right]
		\left[
		-2 C k_x k_{s}^{2} k_z + C^2 k_y k_r^4
		-2 C k_y k_r^2 m +k_y M_z
		\right]
	\end{aligned}
}{
	k_{s}^{4}
	\left[
	k_z^2+\left(-C k_r^2 +m\right)^2
	\right]^2
}\ ,\\
g_{xz}
&=&
\frac{
	\left[C\left(k_{s}^{2}-k_z^2\right)-m\right]
	\left[
	C^2 k_y k_r^4 + C\left(
	-2k_x\left(k_x^2+k_y^2\right)k_z -2 k_y k_r^2 m
	\right)
	+k_y M_z
	\right]
}{
	k_{s}^{2}
	\left[
	k_z^2+\left(-C k_r^2 + m\right)^2
	\right]^2
}\ ,\\
g_{zz}
&=&
\frac{\left[m-C\left(k_s^2-k_z^2\right)\right]^2}
{\left[\left(m-C k_r^2\right)^2+k_z^2\right]^2}\ .
\end{eqnarray}
where, we define $k_{r}^{2} = k_s^{2} + k_z^{2}$, $k_{s}^{2} = k_x^{2} + k_y^{2}$, and $M_z=(k_z^2+m^2)$ to compactly express the above expressions.

\vspace{0.3cm}
\begin{figure*}[h]
	\subfigure{\includegraphics[width=0.75\textwidth]{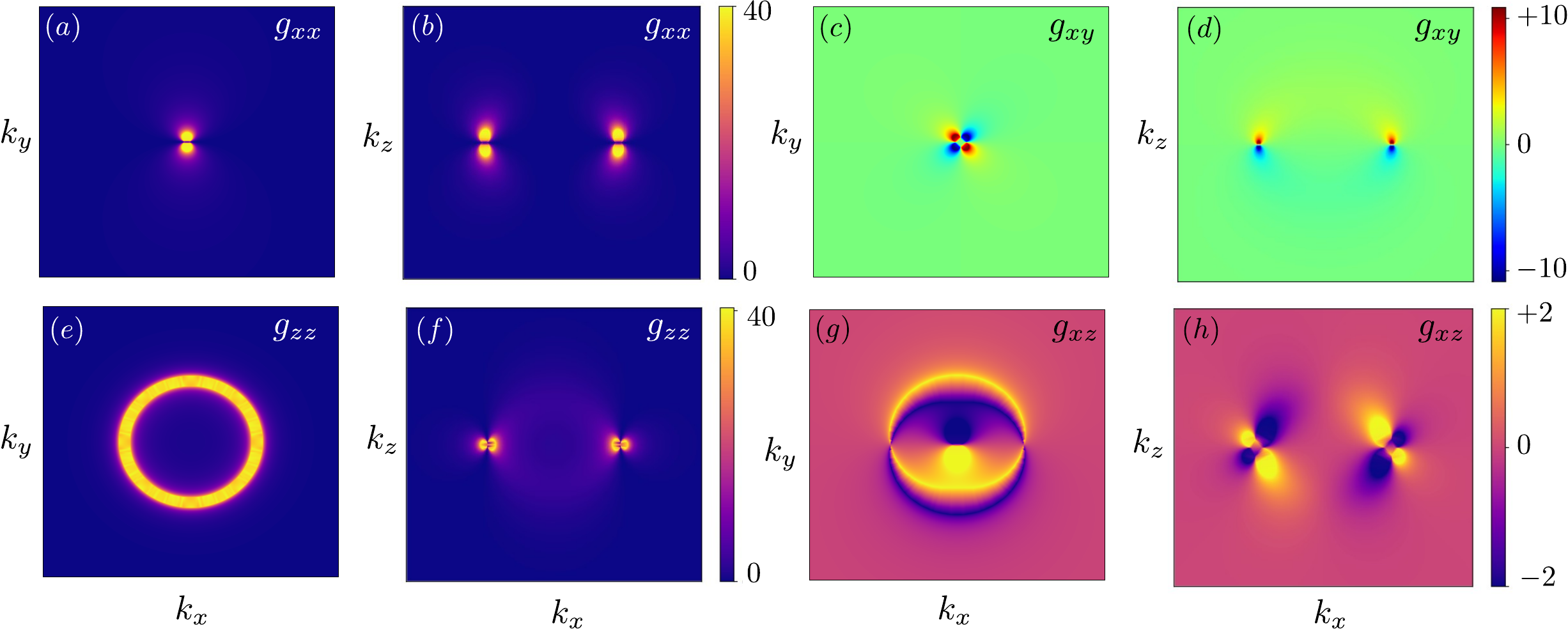}}
	\caption{Different components of the quantum metric are illustrated in various planes of crystal momentum. Panels (a) and (b) show the $xx$-component, $g_{xx}$, in the $k_x$-$k_y$ and $k_x$-$k_z$ planes, respectively. On the other hand, panels (c) and (d) depict the $xy$-component, $g_{xy}$, in the corresponding planes. In panels (e) and (f), we display the $zz$-component, $g_{zz}$, in the $k_x$-$k_y$ and $k_x$-$k_z$ planes, respectively. Finally, panels (g) and (h) showcase 
	the $zx$-component, $g_{zx}$, in the same respective momentum planes. For all the panels, we choose the mass parameter as $m=0.5$.}
	\label{QMs}
\end{figure*}

In Figs.~\ref{QMs}(a) and (b), we show the $xx$-component of the quantum metric, $g_{xx}$, in the $k_x$-$k_y$ and $k_x$-$k_z$ planes, respectively. Similarly, Figs.~\ref{QMs}(c) and (d) present the $xy$ component, $g_{xy}$, in the corresponding momentum planes. Unlike the $xx$-component, which remains predominantly positive, the $xy$-component exhibits both positive and negative extrema due to its off-diagonal nature. Furthermore, we compute and display the remaining components, namely the $zz$ and $zx$ components of the quantum metric, $g_{zz}$ and $g_{zx}$, in the $k_x$-$k_y$ and $k_x$-$k_z$ planes. The corresponding results are shown in Figs.~\ref{QMs}(e), (f), and (g), (h), respectively. Note that, all the components indicate that the Hopf-link semimetal is in the linked phase.


\end{onecolumngrid}	
\end{document}